\documentclass[11pt]{article}
\pdfoutput = 1

\usepackage[utf8]{inputenc}
\usepackage{color,graphicx}
\usepackage{epsfig}
\usepackage{amsmath}
\usepackage{verbatim}
\usepackage{amssymb}
\usepackage{mathabx}
\usepackage[mathcal]{eucal}
\usepackage{stmaryrd}
\usepackage{empheq}
\usepackage{esint}
\usepackage{physics}
\usepackage{amsfonts}
\usepackage{cite}
\usepackage{array}
\usepackage{setspace}
\usepackage{braket}
\usepackage{float}
\usepackage{url}
\usepackage{mathtools}
\usepackage{tensor}
\usepackage{bbold}
\usepackage{slashed}
\usepackage{tikz}
\usepackage{graphicx}
\usepackage{epstopdf}
\usepackage{subfigure}
\usepackage[labelfont=bf,font={sf}]{caption}
\usepackage{pgfplots}
\usepackage{mathrsfs}
\usepackage{fancybox}
\usepackage{eurosym}
\usepackage{tcolorbox}
\usepackage{tensor}
\allowdisplaybreaks

\usepackage[margin = 2.2cm]{geometry}
\usepackage[ragged]{footmisc}
\usepackage[bookmarks=true,bookmarksnumbered=false,
hyperindex=true,bookmarksopen=true,hyperfigures=true,
colorlinks=true,linkcolor=webblue,citecolor=webgreen,urlcolor=webblue,breaklinks]{hyperref}
\definecolor{webgreen}{rgb}{0, 0.5, 0}
\definecolor{webblue}{rgb}{0, 0, 0.5}
\definecolor{webred}{rgb}{0.5, 0, 0}
\definecolor{darkgreen}{rgb}{0,0.5,0}

\newcommand{\average}[1]{\left\langle #1 \right\rangle}

\newcommand{\Uqsu}{U_q(\mathfrak{su}(1,1))}

\newcommand{\Uqsl}{U_q(\mathfrak{sl}(2,\mathbb{R}))}

\newcommand{\dpi}{\mathcal{D}}

\def\ben{\begin{equation}}
\def\een{\end{equation}}

\let\a=\alpha \let\b=\beta \let\g=\gamma \let\d=\delta 
   
     \let\r=v
 \let\t=\tau

 \let\P=\Phi

\def\be{\begin{equation}}
\def\ee{\end{equation}}
\def\ba{\begin{array}}
\def\ea{\end{array}}

\def\dalemb#1#2{{\vbox{\hrule height .#2pt
       \hbox{\vrule width.#2pt height#1pt \kern#1pt
               \vrule width.#2pt}
       \hrule height.#2pt}}}

\newcommand{\bea}{\begin{eqnarray}}
\newcommand{\eea}{\end{eqnarray}}

\def\bfb{{\textbf{\{}}}
\def\bfk{{\textbf{\}}}}

\renewcommand{\d}{\mathrm{d}}
\renewcommand{\i}{\mathrm{i}}

\numberwithin{equation}{section}

\begin{document}

\thispagestyle{empty}
\begin{center}
    ~\vspace{5mm}

     {\LARGE \bf 
    
   The q-Schwarzian and Liouville gravity}
    
   \vspace{0.4in}
    
    {\bf Andreas Blommaert$^1$, Thomas G. Mertens$^2$, Shunyu Yao$^3$}

    \vspace{0.4in}
    {$^1$SISSA and INFN, Via Bonomea 265, 34127 Trieste, Italy\\
    $^2$Department of Physics and Astronomy\\
Ghent University, Krijgslaan, 281-S9, 9000 Gent, Belgium\\
    $^3$Department of Physics, Stanford University, Stanford, CA 94305, USA}
    \vspace{0.1in}
    
    {\tt ablommae@sissa.it, thomas.mertens@ugent.be, shunyu.yao.physics@gmail.com}
\end{center}

\vspace{0.4in}

\begin{abstract}
\noindent We present a new holographic duality between q-Schwarzian quantum mechanics and Liouville gravity. The q-Schwarzian is a one parameter deformation of the Schwarzian, which is dual to JT gravity and describes the low energy sector of SYK. We show that the q-Schwarzian in turn is dual to sinh dilaton gravity. This one parameter deformation of JT gravity can be rewritten as Liouville gravity. We match the thermodynamics and classical two point function between q-Schwarzian and Liouville gravity. We further prove the duality on the quantum level by rewriting sinh dilaton gravity as a topological gauge theory, and showing that the latter equals the q-Schwarzian. As the q-Schwarzian can be quantized exactly, this duality can be viewed as an exact solution of sinh dilaton gravity on the disk topology. For real q, this q-Schwarzian corresponds to double-scaled SYK and is dual to a sine dilaton gravity.
\end{abstract}

\pagebreak
\setcounter{page}{1}
\tableofcontents

\section{Introduction}
Tractable holographic dualities between simple models of low-dimensional quantum gravity have proven to be of paramount importance in our quest for understanding quantum gravity. One such model which has played a central role in recent years is Jackiw-Teitelboim (JT) gravity \cite{jackiw1985lower,teitelboim1983gravitation}. It is a 2d dilaton gravity with a linear dilaton potential $V(\Phi)=2\Phi$ and Euclidean action
\begin{align}
\label{dilgrav}
S = -\frac{1}{2}\int \d x \sqrt{g}\,(\Phi R + V(\Phi))-\int\d\tau \sqrt{h}\Phi K \, .
\end{align}
This is topological in the bulk, but with suitable boundary conditions it can describe boundary graviton fluctuations whose dynamics are governed by the 1d Schwarzian action \cite{Maldacena:2016upp,Engelsoy:2016xyb,Jensen:2016pah}. At lowest order in the topological expansion, this Schwarzian boundary action captures all physics of the gravitational bulk, and in fact JT gravity is precisely dual to the Schwarzian at the quantum level.
Comparing different dilaton gravity models \eqref{dilgrav}, the JT model is unique in the sense that the bulk field $\Phi$ is a Lagrange multiplier which can hence be path-integrated out, leading directly to a pure-boundary description of the reduced dynamics. This is no longer true for generic dilaton gravity models. Nonetheless, the bulk is topological in all cases so a pure-boundary description should be possible. In this paper, we'll work out such a holographic duality, for one specific potential (although our methods in principle apply to generic potentials).

In \cite{Blommaert:2023opb}, we studied the double-scaled version of the SYK model \cite{Cotler:2016fpe,berkooz2018chord,berkooz2019towards} and found a description in terms of a simple 1d quantum mechancal system, which we referred to as the q-Schwarzian. Here $0<q<1$ is related to the double-scaling parameter as $q=e^{-p^2/N}$, where $N$ is the number of distinct fermions, and $p$ is the number of fermions participating in any single interaction. This parameter $q$ is technically the deformation of the quantum group SU$_q(1,1)$ as one deforms away from the JT gravity description at $q=1$, governed by the classical group SU$(1,1)$. We argued for a bulk dilaton gravity description in terms of a sine dilaton potential (see also \cite{Goel:2022pcu}), postponing some of the details to later work. Here we'll study a related theory, interesting in its own right, which shares many of the features of the DSSYK story but has slightly fewer complications. We'll briefly comment on the new ingredients that enter for the DSSYK case in section \ref{sect:dssyk}.

In this work we focus on the same q-Schwarzian theory, but now with $\abs{q}=1$ which is  conveniently parametrized as
\begin{equation}
    q=e^{\pi i b^2}\,,\quad b \in \mathbb{R}\,.\label{1.2}
\end{equation}
This corresponds to a sinh dilaton gravity
\begin{equation}
    \boxed{V_\text{qsch}(\Phi)=\frac{\sinh(2\pi b^2\P)}{\pi b^2}\,}\,.\label{1.3}
\end{equation}
This particular dilaton gravity has attracted quite some attention recently, and is referred to as Liouville gravity \cite{Mertens:2020hbs,Fan:2021bwt,Kyono:2017pxs,Suzuki:2021zbe,Goel:2020yxl,Collier:2023cyw}. The goal of this work is to argue that the q-Schwarzian is the precise holographic dual (or boundary description) of Liouville gravity (or sinh dilaton gravity) at disk level. 

Our work is structured as follows
\begin{enumerate}
    \item In \textbf{section \ref{Sec.qsch}} we study the classical solutions (subsection \ref{sect:2.1}), and symmetry algebra (subsection \ref{sect:2.2}) of the q-Schwarzian and match these to the classical limits of the known bulk descriptions of Liouville gravity and sinh dilaton gravity. 
    \item In \textbf{section \ref{sect:PIargument}} we prove that the theories are exactly dual by formulating the bulk theory in terms of a topological Poisson-sigma model \cite{Ikeda:1993aj, Ikeda:1993fh, Schaller:1994es}. For the latter, we'll arive at a boundary description that reduces to the q-Schwarzian in our case of interest.
    \item In \textbf{section \ref{sect:conc}} we briefly discuss the case of DSSYK and some interesting additional features which arise in its gravitational dual. Furthermore we discuss quantization of the q-Schwarzian and show that this reproduces the exact amplitudes of Liouville gravity \cite{Mertens:2020hbs}. In combination with the results of section \ref{sect:PIargument} one can view this as an exact solution of sinh dilaton gravity,\footnote{This same logic was essentially used to exaclty solve JT gravity in \cite{Blommaert:2018oro,Iliesiu:2019xuh}.} and one can interpret our results as a (quantum) proof that the latter equals Liouville gravity.
\end{enumerate}
Next we introduce both sides of the duality (Liouville gravity and the q-Schwarzian) in more detail.
\subsection{Liouville gravity}
Liouville gravity \cite{polyakov1981quantum,david1988conformal,distler1989conformal} is a non-critical string combining a 2d Liouville CFT, a matter CFT, and the $bc$-ghost CFT such that the total central charge vanishes. We will consider the worldsheet to have disk topology. For our purposes, we can take the matter sector to be a timelike Liouville field $\chi$, and central charge ($b\in \mathbb{R}$ parameterizes the model)
\begin{equation}
    c_\chi=1-6(b-1/b)^2\,.
\end{equation}
Upon fixing the worldsheet metric to conformal gauge with scale factor $e^{2 b \varphi}$ one finds a second Liouville action \cite{polyakov1981quantum,david1988conformal,distler1989conformal} for $\varphi$ with central charge
\begin{equation}
    c_\varphi=1+6(b+1/b)^2\,.
\end{equation}
Upon doing the field redefinition
\begin{equation}
    \varphi=\frac{\rho}{b}-b\pi\Phi\,,\quad \chi=\frac{\rho}{b}+b\pi\Phi\, ,
\end{equation}
the action of the total model reduces to sinh dilaton gravity \eqref{3.6}, with $\rho$ the conformal factor of the spacetime metric. Even though several points of the relation between Liouville gravity and sinh dilaton gravity are not perfectly understood, this particular step is well-documented \cite{Mertens:2020hbs,Fan:2021bwt,StanfordSeiberg,Suzuki:2021zbe,Collier:2023cyw,Goel:2020yxl,Turiaci:2020fjj}. Disk amplitudes with a fixed boundary length $\ell$ can be found \cite{Mertens:2020hbs} by Laplace transforming those that have FZZT boundary conditions \cite{Fateev:2000ik,Teschner:2001rv} ($\sim$ fixed energy) on the $\varphi$ field
\begin{equation}
    \mu_B\sim \cosh(\a)\,,
\end{equation}
and ZZ boundary conditions for the $\chi$ field.\footnote{The structure constants associated with $\chi$ have no $\alpha$ dependence (as they satisfy ZZ, not FZZT boundary conditions) thus they do not show up in equations such as \eqref{3.22}, where one Laplace transforms with respect to $\mu_B$. Other combinations of boundary conditions could be interesting, such as matching energies $\alpha_\phi=\alpha_\chi$ (see e.g. \cite{Narovlansky:2023lfz,Suzuki:2021zbe,Collier:2023cyw}). But this is not what was studied in \cite{Mertens:2020hbs}. It would be interesting to know if other combinations have q-Schwarzian interpretations.\label{fn:zz}} 

Much like DSSYK amplitudes have an interpretation in terms of SU$_q(1,1)$ representation theory \cite{Berkooz:2022mfk,Blommaert:2023opb}, Liouville gravity amplitudes have an underlying quantum group structure \cite{Fan:2021bwt,Mertens:2022aou}: the modular double of SL$_q(2,\mathbb{R})$ \cite{Faddeev:1999fe,Ponsot:1999uf,Ponsot:2000mt,Kharchev:2001rs,Bytsko:2002br,Bytsko:2006ut}, the latter being a different q-deformation of the classical group SU$(1,1)$ \cite{klimyk2012quantum}.

\subsection{The q-Schwarzian}
The q-Schwarzian theory \cite{Blommaert:2023opb} is defined by the Euclidean action
\begin{align}
  \boxed{ S_\text{qsch} = -\int_0^\beta \d u \bigg(\i \sum_A p_A x_A'+\frac{\cosh(\pi b^2 p_\varphi)}{2\pi^2b^4}-\frac{\mu_\b\mu_\g}{\pi^2b^4}e^{-2\varphi+ \pi b^2 p_\varphi} \bigg)}\,, \label{def:qsch}
\end{align}
with coordinates $x_A=(\varphi,\beta,\gamma)$ and associated momenta $p_A=(p_\varphi,p_\beta,p_\gamma)$.\footnote{ Throughout we put $\hbar=1$ but we do remark that $\hbar$ plays a role in explaining the difference between the classical Poisson brackets of the currents \eqref{algebra} and its quantized version \eqref{eq:qalg} with $b^2_\text{quant}=\hbar\, b^2_\text{class}$. This is explained in more detail in section 4 of \cite{Blommaert:2023opb}. We checked that putting $\hbar$ back everywhere leads to consistent results.\label{fn1}
} We introduced the shorthand notation:
\begin{equation}
    \mu_\b=\frac{e^{2\pi b^2 \b  p_\b}-1}{-2\i\b}\,,\quad \mu_\g=\frac{e^{2\pi b^2 \g p_\g}-1}{-2\i\g}\,.
\end{equation}
In section \ref{sect:PIargument} we will use a Lorentzian action where $\beta= -\i T$
\begin{align}
   S_\text{qsch} = \int_0^T \d t \bigg(\sum_A p_A x_A'-\mathbf{H}(x_A,p_A)\bigg)\,,\quad \mathbf{H}(x_A,p_A)=\frac{\cosh(\pi b^2 p_\varphi)}{2\pi^2b^4}-\frac{\mu_\b\mu_\g}{\pi^2b^4}e^{-2\varphi+\pi b^2 p_\varphi}\,. \label{6dimqsch}
\end{align}
In section \ref{Sec.qsch} it will be convenient to work with rescaled variables in \eqref{def:qsch}
\begin{equation}
    \frac{\d u}{\pi b^2}= \d \t\, ,\quad \pi_A= \i \pi b^2 p_A\,,\label{rescale}
\end{equation}
resulting in a Euclidean action which behaves manifestly semiclassical for $b\to 0$
\begin{equation}
 \begin{aligned}
       S_{\text{qsch}} = -\frac{1}{\pi b^2}\int_0^{\beta/\pi b^2}\d \tau \bigg(\pi_\varphi \varphi'+\pi_\beta \beta'+\pi_\g \g'+\frac{1}{2}\cos(\pi_\varphi)-\mu_\b \mu_\g e^{-2\varphi-i\pi_\varphi}\bigg)\,\, .
    \end{aligned}
    \label{eq:ac}
\end{equation}
Throughout this work, we will use the term q-Schwarzian as a moniker for three theories. The first is \eqref{def:qsch} and describes quantum mechanics on SL$^+_q(2,\mathbb{R})$, see section \ref{sect:2.2} for motivation of this name. The second theory is \eqref{def:qsch} with one constraint \eqref{oneconstraint}. The constraint leads to a 4d phase space path integral which we referred to as q-Schwarzian in \cite{Blommaert:2023opb}. This is the theory which for $b\to 0$ (and low energies) reduces to the ordinary Schwarzian model, see e.g. section 2.1 in \cite{Blommaert:2023opb}.
In a holographic context this is relevant for a ``one-sided setup'' describing a Rindler-time slicing. This is the setup we'll consider in section \ref{sect:PIargument}. 
Thirdly, there is \eqref{def:qsch} with \emph{two} constraints \eqref{twoconstraint}, resulting in a 2d phase space which we denoted q-Liouville in \cite{Blommaert:2023opb}. Holographically this is relevant for a global slicing, and referred to as ``two-sided''. This is the setup in which quantization is easiest, and it is also simplest to have in mind when thinking about the classical solutions in section \ref{sect:2.1}.\footnote{The classical solutions are relevant to all cases, including the one-sided q-Schwarzian, but the interpretation of $e^{-2\Delta\varphi}$ as boundary two-point function later on in subsection \ref{sect:3.2} only applies in the q-Liouville formulation. For other applications of the Liouville formulation of JT gravity see for instance \cite{Bagrets:2016cdf,Harlow:2018tqv}.} We can summarize the slicings of the Euclidean disk (along with several other details that will be relevant later) for the two relevant theories graphically as follows:
\begin{equation}
    \begin{tikzpicture}[baseline={([yshift=-.5ex]current bounding box.center)}, scale=0.7]
 \pgftext{\includegraphics[scale=1]{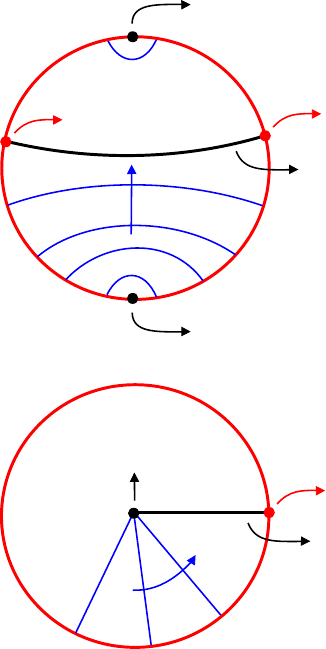}} at (0,0);
    \draw (3,5.4) node {$\varphi(\beta)=-\infty$ \eqref{2.19}};
    \draw (3,-0.2) node {$\varphi(0)=-\infty$ \eqref{2.19}};
    \draw (4.4,-0.9) node {smooth boundary initial state};
    \draw (-0.7,3.4) node {\color{red}\eqref{twoconstraint}};
    \draw (4.95,3.5) node {\color{red}\eqref{oneconstraint} constraint};
    \draw (4.1,2.6) node {Cauchy slice};
    \draw (4.1,1.9) node {2d phase space};
    \draw (5,-2.85) node {\color{red}\eqref{oneconstraint} constraint};
    \draw (4.25,-3.7) node {Cauchy slice};
    \draw (4.25,-4.4) node {4d phase space};
    \draw (2.1,-5.4) node {\color{blue}time flow};
    \draw (-6,3.8) node {two sided};
    \draw (-6,3) node {\textbf{q-Liouville}};
    \draw (-6,2.3) node {section \ref{Sec.qsch}};
    \draw (-6,1.6) node {section \ref{sect:quant}};
    \draw (-6,-2.2) node {one sided};
    \draw (-6,-3) node {\textbf{q-Schwarzian}};
    \draw (-6,-3.7) node {section \ref{sect:PIargument}};
    \draw (-0.3,-2.05) node {free label};
    \draw (2.5,0.85) node {\color{blue}time flow};
  \end{tikzpicture}\label{2.6}
\end{equation}
The constraints \eqref{oneconstraint} and \eqref{twoconstraint} implement the analogue of Brown-Henneaux \cite{brown1986central} holographic boundary conditions \eqref{constraint} in the gravitational dual, as we detail in section \ref{sect:com}.

\textbf{Comment about temperature.} We will be interested in values $\beta$ chosen so that a rescaled version of the temperature remains finite upon taking $b\to 0$
\begin{equation}
    \frac{\beta}{\pi b^2}=\text{ finite}\,.\label{1.13}
\end{equation}
In the q-Schwarzian we chose our definition of $\beta$ by the choice in \cite{Blommaert:2023opb} to set the Hamiltonian equal to the Casimir of the quantum group $\mathbf{H}=L_C$.\footnote{In the context of DSSYK, this corresponds to keeping $\beta_\text{DSSYK} J$ finite when $q\to 1$, which is the large $p$ limit\cite{Maldacena:2016hyu,Lin:2023trc}. Our choice of normalization is different from the normalization of the Hamiltonian in DSSYK \cite{berkooz2019towards,Lin:2022rbf} by the factor mentioned above.} Similarly in dilaton gravity we have a choice on what we call the physical boundary time. One can therefore easily modify our conventions $\beta/\pi b^2\to \beta$ if desired. In the rescaled action \eqref{eq:ac} we indeed see that finite $\beta/\pi b^2$ leads to a reliable saddle for $b\to 0$.
\section{The q-Schwarzian}\label{Sec.qsch}
The main object of study in this work is the path integral associated to the action \eqref{def:qsch}. In this section, we'll look at the saddle of this path integral and compare it to the saddle of both the Liouville gravity path integral and the sinh dilaton gravity path integral.

\subsection{Classical solutions}\label{sect:2.1}
For technical convenience, we consider the path integral \eqref{def:qsch} in rescaled variables \eqref{eq:ac}
\begin{equation}
\begin{aligned}
        Z_\text{qsch}(\beta)=\int \dpi & \varphi \dpi \pi_\varphi\dpi \beta \dpi\pi_\beta \dpi \g \dpi \pi_\g\\&\exp\bigg(\frac{1}{\pi b^2}\int_0^{\beta/\pi b^2}\d \tau \bigg(\pi_\varphi \varphi'+\pi_\beta \beta'+\pi_\g \g'+\frac{1}{2}\cos(\pi_\varphi)-\mu_\b \mu_\g e^{-2\varphi-i\pi_\varphi}\bigg) \bigg)\,.
    \end{aligned}\label{rescaledac}
\end{equation}
We first discuss the classical solutions. The classical saddles dominate the path integral when $b\to 0$, keeping in mind that we also keep $\beta/\pi b^2$ fixed \eqref{1.13}. 

We first consider the Hamilton equations for $\pi_\b$ respectively $\b$
\begin{equation}
    \beta'=\mu_\g e^{-2\varphi-\i\pi_\varphi}e^{-2\i\b\pi_\b}\,,\quad \pi_\beta'=-\mu_\g e^{-2\varphi-\i\pi_\varphi}\frac{\d \mu_\b}{\d\b}\,,
\end{equation}
from which one deduces that $\mu_\beta$ is a constant of the motion
\begin{equation}
\mu_\b'=e^{-2\i\b\pi_\b}\pi_\beta'+\frac{\d \mu_\b}{\d\b}\beta'=0\,.
\end{equation}
This trivially extends to $\mu_\g$. With this knowledge one can directly solve the Hamilton equations for $\pi_\varphi$ respectively $\varphi$, for fixed constants (characterizing the solution) $\mu_\b,\mu_\g$
\begin{align}\label{EOMsch}
    &\varphi'=\frac{1}{2}\sin(\pi_\varphi)-\i\, \mu_\g \mu_\b e^{-2\varphi-\i\, \pi_\varphi}\,,\quad \pi_\varphi'=2\mu_\g \mu_\b e^{-2\varphi-\i\, \pi_\varphi}\,.
\end{align}
For future reference we notice that (using the fact that $\mu_\b$ is constant) one finds
\begin{equation}
    (\b\pi_\b)'=\mu_\g \mu_\b e^{-2\varphi-\i\, \pi_\varphi}\,.
\end{equation}
Among other things, one immediately finds from this another constant of motion $\b\pi_\b-\pi_\varphi/2$. In total, since we have a 6d phase space, there will be 6 such constants. We discuss their physical relevance later in section \ref{sect:2.2}. Taking the derivative of the last equation in \eqref{EOMsch} one finds
\begin{equation}
    \varphi'+\frac{1}{2}\i\, \pi_\varphi'=-\frac{1}{2}\frac{\pi_\varphi''}{\pi_\varphi'}\,,\label{2.11}
\end{equation}
which upon insertion in the first equation of \eqref{EOMsch} gives
\begin{equation}
    \sin(\pi_\varphi)=-\frac{\pi_\varphi''}{\pi_\varphi'}\,.\label{2.12}
\end{equation}
The solution of this equation is parametrized by two integration constants, $\alpha$ and $d$,
\begin{align}
   \tan(\pi_\varphi/2)=\coth(\a/2)\tan(\sinh (\a)\t/2+d)\,,\quad 4\mu_\b\mu_\g\,e^{-2\varphi}=-\frac{\sinh(\a)^2}{\sin^2(\sinh (\a)\t/2+d+\i\a/2)}\,.\label{2.13}
\end{align}
The (conserved) energy of the solution follows from the Hamiltonian term in \eqref{rescaledac} and can be evaluated explicitly using \eqref{2.13}
\begin{equation}
    \pi^2 b^4 E=-\frac{1}{2}\cos (\pi_\varphi)+\mu_\b \mu_\g\,e^{-2\varphi-\i\pi_\varphi} =-\frac{1}{2}\cos (\pi_\varphi)+\frac{1}{2}\varphi'=\frac{1}{2}\cosh(\a)\,.\label{enenen}
\end{equation}
This identifies $\a$ as (parameterizing) the energy of the classical solutions. Simultaneously, we note that the solutions \eqref{2.13} have periodicity
\begin{equation}
    0<\tau<\frac{2\pi}{\sinh(\a)}\,,
\end{equation}
which determines the relation between inverse temperature and $\alpha$. From this one determines their entropy using $\beta=\d S/\d E$ such that $\pi b^2S=\pi\a\,+\,$constant. Alternatively we can immediately compute this as the area in phase space enclosed by the classical orbit\footnote{We used the fact that $\mu_\b$ is constant to solve for $\pi_\b(\b,\mu_\b)$ such that $\pi_\b(\b,\mu_\b)\b'$ is recognized as a total derivative, and ditto for $\pi_\g \g'$. We substituted furthermore $\varphi'$ using \eqref{2.11} and \eqref{2.12}, recognized one total derivative, and used integration by parts. Because $\pi_\varphi'$ is reasonably simple the resulting integral can be numerically evaluated for generic complex $d$.}
\begin{equation}
  \pi b^2S=  \int_0^{2\pi/\sinh(\a)}\d \tau \bigg(\pi_\varphi \varphi'+\pi_\beta \beta'+\pi_\g \g'\bigg)=\frac{1}{2}\int_0^{2\pi/\sinh(\alpha)}\d \t \,\pi_\varphi' \log (\pi_\varphi')=\a\pi-\pi\log(2)\,.
\end{equation}
Such relations between the classical phase space trajectory and the density of states $e^S$ are key elements in the periodic orbit literature, see chapter 9 of \cite{Haake:1315494}. The partition function takes the form
\begin{equation}
    Z_\text{qsch}(\beta)=\int_0^\infty \d E\, e^{S(E)-\beta E}\,,
\end{equation}
so on the classical saddle one can either evaluate $S$ using the energy saddle $\beta=\d S/\d E$ or by evaluating directly the phase space volume at fixed energy (the on-shell action without the Hamiltonian term). In any case the classical approximation of the q-Schwarzian partition function \eqref{rescaledac} is
\begin{equation}
    \boxed{Z_\text{qsch}(\beta)_\text{class}\approx  \int_0^\infty \d \a \sinh(\a) \exp\bigg(-\frac{1}{b^2}\log(2)+ \frac{\a}{b^2}-\beta \frac{\cosh(\a )}{2\pi^2 b^4}\bigg)}\,.\label{onshellqsch}
\end{equation}
We'll match this with the partition function of Liouville gravity and sinh dilaton gravity in section \ref{sect:thermo}.

\textbf{Comment about approximations.} Our goal is to reproduce the on-shell action to leading order in $1/b^2$, this is what we'll mean by the $\approx$ symbol. We remark that the integration measure $\sinh(\a)$ in this sense contributes at the one-loop level. Although natural, it should not be viewed as a trustworthy prediction of our q-Schwarzian analysis, because we did not attempt a one-loop calculation. Let us also remind the reader that we choose $\beta$ such that $\beta/\pi b^2$ is order one \eqref{1.13}, such that indeed all terms in the action in \eqref{onshellqsch} are order $1/b^2$. This comment extends to all on-shell analyses below.

Before proceeding, we should make two comments about integration constants. The complex time-shift parameter $d$ can be fixed by imposing
\begin{equation}
    \varphi(0)=-\infty\,.\label{2.19}
\end{equation}
Here we have in mind the two-sided q-Liouville version of the q-Schwarzian (as explained around \eqref{2.6}). This constraint follows in the two-sided language from imposing ``smoothness'' at $\tau=0$, as explained in the paragraph below (4.20) in \cite{Blommaert:2018oro} for JT. In the quantization language \cite{Fan:2021bwt,Blommaert:2023opb} (and section \ref{sect:quant}) this corresponds with starting from the identify (quantum) group element, taking into account the change of basis. This fixes $d=-\i\a/2$. Alternatively, the reader may view this as part of the definition of the q-Liouville path integral (which we design to match with Liouville gravity amplitudes).

Secondly, the q-Liouville system comes with two constraints \eqref{twoconstraint} on phase space (Brown-Henneaux boundary conditions, see discussion around \eqref{2.6} and section \ref{sect:vielbein}) which imply
\begin{equation}
    4\mu_\b\mu_\g=1\,.\label{2.20}
\end{equation}
Summarizing, the version of the classical solution \eqref{2.13} relevant for the q-Liouville setup is
\begin{equation}
    \boxed{
    e^{-2\varphi}=-\frac{\sinh(\a)^2}{\sin^2(\sinh (\a)\t/2)}
    }\,.\label{sch2ptans}
\end{equation}
It is this expression which naturally has the interpretation as a boundary-to-boundary two-point function (see also section \ref{sect:5.3}), and which we match with the Liouville gravity two-point correlator in section \ref{sect:3.2}.

\subsection{Matching with Liouville gravity - Thermodynamics} \label{sect:thermo}
We now want to match \eqref{onshellqsch} with the semiclassical partition function of Liouville gravity (the latter was reported in \cite{Mertens:2020hbs}, but we present it in a slightly different fashion here). The (disk) partition function of Liouville gravity was computed in \cite{Fateev:2000ik,Seiberg:2003nm} and phrased in a suggestive language in \cite{Saad:2019lba,Mertens:2020hbs} as\footnote{For concreteness the boundary conditions we consider are
\begin{equation}
     \ell_{12}=\int_{x_1}^{x_2}\d x\, e^{b\varphi}=\frac{\beta_{1 2}}{2\pi b^2}\,.\label{3.21}
\end{equation}
Here $x_i$ are worldsheet moduli to be integrated over but the $\beta_{i j}$ are physical lengths (Euclidean times to be more precise) on the holographic boundary in the dilaton gravity formulation.}
\begin{equation}
    Z(\beta)=\int_0^\infty \d s\,\frac{1}{S_b(\pm 2\i s)}\,\exp\bigg(-\beta \frac{\cosh(2\pi b s)}{2\pi^2 b^4}\bigg)\,.\label{3.0}
\end{equation}
Here and below we ignore overall constants as those do not affect the saddle point analysis (the classical limit). The double-sine function $S_b(x)$ appears often in the Liouville literature, its definition and several useful properties can be found in \cite{ip2021tensor}. This includes a useful equation for the density of states
\begin{equation}
    \frac{1}{S_b(\pm 2\i s)}=4\sinh(2\pi b s)\sinh(2\pi s/b)\,.\label{3.1}
\end{equation}
Semiclassical physics emerges when the density of states is large, because of the large action the integrals are dominated by saddle point. Here we can achieve this by taking $b\to 0$ while keeping $\a=2\pi b s$ finite. The partition function becomes
\begin{equation}
Z(\beta)_\text{class}\approx \int_0^\infty \d \a \sinh(\a) \exp\bigg( \frac{\a}{b^2}-\beta \frac{\cosh(\a )}{2\pi^2 b^4}\bigg)=\int_0^\infty \d E\, e^{S(E)-\beta E}\,,\label{3.4}
\end{equation}
from which one deduces classical thermodynamic quantities, for instance the energy-temperature relation \cite{Mertens:2020hbs}
\begin{equation}
    \frac{\beta}{\pi b^2}=\frac{2\pi}{\sinh(\a)}\,.\label{3.5}
\end{equation}
This matches the classical limit of the q-Schwarzian path integral \eqref{onshellqsch}. However, we do remark that there is an overall degeneracy $S_0$ mismatch because of the $\log(2)$ in \eqref{onshellqsch}. This was to be expected as we did not track overall normalization, nevertheless it would be interesting to match more precisely.

This partition function can alternatively be computed directly using the sinh dilaton gravity action, where we have the following Euclidean path integral action\footnote{This model made appearances for instance in \cite{Mertens:2020hbs,Fan:2021bwt,StanfordSeiberg,Suzuki:2021zbe,Collier:2023cyw}.}
\begin{equation}
    \boxed{S_\text{grav}=-\frac{1}{2}\int \d x \sqrt{g}\bigg(\Phi R + \frac{\sinh(2\pi b^2\P)}{\pi b^2}\bigg)-\int\d\tau \sqrt{h}\Phi K\,}\,.\label{3.6}
\end{equation}
By rescaling $\pi b^2 \Phi\to \Phi$ the whole action is multiplied by $1/b^2$, so the classical saddle point is trustworthy for $b\to 0$ with the new rescaled $\Phi \sim 1$.\footnote{This is distinct from the JT gravity regime, where we take $b\to 0$ whilst keeping the original $\Phi\sim 1$. This corresponds to zooming in on low energies.} One may gauge-fix the classical solution to \cite{Gegenberg:1994pv,Witten:2020ert}
\begin{equation}
    \d s^2=F(r)\d\t^2+\frac{\d r^2}{F(r)}\,,\quad \Phi=r\,.\label{3.7}
\end{equation}
In this gauge $R=-F''$, and we see that the $\Phi$ EOM in \eqref{3.6} imply $F''=V'$, which for our potential \eqref{1.3} results in the classical metric
\begin{equation}
    F(r)=\frac{\cosh(2\pi b^2 r)}{2\pi^2 b^4}-\frac{\cosh(2\pi b^2 \Phi_h)}{2\pi^2 b^4}\,.\label{3.8}
\end{equation}
This classical spacetime has a black hole horizon at $r=\Phi_h$ with the interior being $r<\Phi_h$. Near $r=\infty$ the metric blows up, making this the natural candidate for a holographic boundary, a suspicion we will confirm in section \ref{sect:com}.

To obtain the on-shell action in the form $S(E)-\beta E$, we want to count classical solutions for which the size of the thermal circle $\beta$ (the range of $\tau$ in \eqref{3.7}) is independent from the Hawking temperature (which is determined by $\Phi_h$ in \eqref{3.8}). Such Euclidean spacetimes have conical singularities at their horizon\footnote{The strength of the conical singularity can be computed using the Gauss-Bonnet theorem:
\begin{equation}
    \frac{1}{4\pi}\int \d x \sqrt{g}R+\frac{1}{2\pi}\int \d \tau \sqrt{h}K\overset{!}{=}1.
\end{equation}}
\begin{equation}
    \sqrt{g}R=-V'+\delta(x-x_h)(4\pi - \beta V(\P_h))\,.\label{3.11}
\end{equation}
For fixed $\beta$ this describes a conical solution for every $\Phi_h$. Then we should integrate over all $\Phi_h$. The rigorous statement is that the conical spacetimes are classical solutions in a context where we first fix the horizon area \cite{Carlip:1993sa,Susskind:1994sm}, in this context $\Phi_h$, and in the end integrate over area \cite{Dong:2018seb,Dong:2022ilf}, with a flat measure. This is exact when inserting the partition function of a disk with a defect \cite{Blommaert:2023opb}. Here we replace the latter by its classical saddle. Via this procedure we obtain the answer
\begin{align}
Z(\beta)_\text{class}&\approx\int_0^\infty \d \Phi_h \exp\bigg(2\pi \Phi_h+\b F(\P_\text{bdy})\bigg)\,.\label{3.12}
\end{align}
Here we inserted \eqref{3.11}, used integration by parts, assumed the holographic boundary of the spacetime lies at fixed $r$, and used $\sqrt{h}K=V/2$. The temperature dependence could have been obtained alternatively by solving the radial WdW equation for 2d dilaton gravity, resulting in the exact result \cite{henneaux1985quantum,Louis-Martinez:1993bge,Iliesiu:2020zld}\footnote{It should be clear that following the same logic as \cite{Iliesiu:2020zld} we can use the exact Liouville gravity partition for the asymptotic case to derive that quantum mechanically $Z(\Phi_h)=\rho(\Phi_h)$ for finite cutoff sinh dilaton gravity.}
\begin{equation}
    Z(\beta)=\int_0^\infty \d \Phi_h\, Z (\Phi_h)\,\exp\bigg(\beta \sqrt{h}\, F(\Phi_\text{bdy})^{1/2}\bigg)\,.
\end{equation}
The fact that the wavefunction $Z(\Phi_h)$ behaves classically as $e^{2\pi \Phi_h}$ is the Bekenstein-Hawking formula. We note that the one-loop measure factor in \eqref{3.4} is not reproduced by this argument, as we did not attempt to evaluate the disk with a defect at one loop.

Notice also that from this equation one reads of the ADM energy as
\begin{equation}
    E=\sqrt{h}\frac{\delta I}{\delta \sqrt{h}}=-F(\Phi_\text{bdy})\,.
\end{equation}
We claim that the holographic boundary for our theory lies at $\Phi_\text{bdy}=\infty$ such that
\begin{equation}
    E=-\infty+\frac{\cosh(2\pi b^2 \Phi_h)}{2\pi^2 b^4}\,.\label{3.15}
\end{equation}
This infinite piece can be stripped off by adding a local counterterm on the boundary, we will ignore such a constant shift for now. (And the correct counterterm is actually suggested by the path integral argument of section \ref{sect:PIargument}, starting from the q-Schwarzian.) Redefining
\begin{equation}
    \alpha=2\pi b^2 \Phi_h\,,
\end{equation}
the gravitational partition function reproduces the q-Schwarzian and Liouville CFT calculations (up to the one loop determinant)
\begin{equation}
    Z(\beta)_\text{class}\approx\int_0^\infty \d \a \,\exp\bigg( \frac{\a}{b^2}-\beta \frac{\cosh(\a )}{2\pi^2 b^4}\bigg)\,.
\end{equation}
We note that this analysis does not reproduce the measure factor in for instance \eqref{oneconstraint}, since we just approximated the defect partition function by its on-shell action.\footnote{In the JT case \cite{Blommaert:2023vbz} indeed the correct measure comes from the one-loop factor of the defect partition function. Defects in sinh dilaton gravity are not well understood. It would be interesting to study them in more detail and match with Liouville results \cite{Mertens:2020hbs}, and reproduce the correct measure here.}

One final comment about the bulk geometry \eqref{3.7} is that (unlike in aAdS$_2$) one does not need renormalization of geodesic distances, for instance the spatial distance between the horizon and the boundary is \emph{finite}.
\subsection{Matching with Liouville gravity - Classical two point function}\label{sect:3.2}
A second observable for which we want to compare the semiclassical answer in Liouville gravity with the q-Schwarzian classical saddle is the boundary two-point function \eqref{sch2ptans}. The natural candidate to attempt to match with are the so-called open string tachyon vertex operators \cite{knizhnik1988fractal}:\footnote{These are the natural operators in Liouville gravity. They are inserted on the boundary as open string insertions. This is one of the points where the combinations of ZZ and FZZT boundary conditions that we choose here will turn out crucial to match with the q-Schwarzian, see also footnote \ref{fn:zz}. Perhaps with different boundary conditions one could match Liouville boundary correlators with different observables in the q-Schwarzian \cite{Narovlansky:2023lfz}.}
\begin{equation}
    \mathcal{B}_{\beta_M}\sim \int_{\partial\Sigma}\d x\,e^{\beta_M\chi }e^{(b-\beta_M)\varphi},
\end{equation}
obtained by gravitationally dressing a matter CFT (timelike Liouville) operator, and integrating this over the boundary to make it diff-invariant. Here we are interested in a semiclassical approximation, where these probes do not backreact on the geometry. This suggest to consider $\beta_M=b\Delta$ with $\Delta$ not scaling with $b\to 0$. We propose the dictionary
\begin{equation}
    \average{\mathcal{B}_{b \Delta}\mathcal{B}_{b\Delta}}_{\tau,\,\beta-\tau} =\langle e^{-2\Delta \varphi(\tau)}\rangle_\beta \,,\label{dic}
\end{equation}
where the boundary vertex operators are separated by \emph{fixed-length} boundary segments of lengths $\tau$ and $\beta-\tau$ respectively \eqref{3.21}. With minor imagination one can depict the relevant correlators in both theories in the same way (to be compared to \eqref{2.6}, where now the black line denotes the insertion of $e^{-2\Delta\varphi}$ at fixed time \eqref{4.11bis}):
\begin{equation}
\label{fig:opslice}
    \begin{tikzpicture}[baseline={([yshift=-.5ex]current bounding box.center)}, scale=0.7]
 \pgftext{\includegraphics[scale=1]{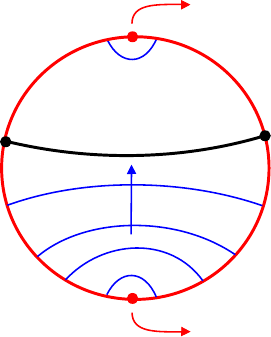}} at (0,0);
    \draw (3.4,2.8) node {\color{red}$\varphi(\beta)=-\infty$ \eqref{2.19}};
    \draw (3.4,-2.8) node {\color{red}$\varphi(0)=-\infty$ \eqref{2.19}};
    \draw (4.8,-3.5) node {\color{red}smooth boundary initial state};
    \draw (3.1,0.6) node {$\mathcal{B}_{b \Delta}$};
    \draw (-3.1,0.6) node {$\mathcal{B}_{b \Delta}$};
    \draw (-2.4,-1.3) node {$\tau$};
    \draw (2.75,1.5) node {$\beta-\tau$};
  \end{tikzpicture}
\end{equation}
Such fixed-length boundary correlators where discussed in \cite{Mertens:2020hbs,Fan:2021bwt} to which we refer the reader for more background. The LHS is evaluated in Liouville gravity, whereas the RHS is a boundary q-Schwarzian amplitude. We do not have a first-principles argument for this specific operator dictionary, instead presenting it as an observation and leaving the underlying reason as an open question.\footnote{As motivation, it is well-known that \eqref{dic} is true in the JT limit where $s=b k$ for $b\to 0$, see equation (4.44) in \cite{Mertens:2020hbs}. In the JT case $2\varphi=\ell$ the (renormalized) geodesic length between operator insertions on the holographic boundary \cite{Blommaert:2018oro,Yang:2018gdb,Saad:2019pqd,Harlow:2018tqv,Lin:2022zxd}.}

Labeling the energies as $\alpha \equiv 2 \pi b s$ as before, the fixed-length Liouville gravity two point function in question was computed as \cite{Mertens:2020hbs}
\begin{align}
    \average{\mathcal{B}_{b \Delta}\mathcal{B}_{b\Delta}}_{\tau,\,\beta-\tau}&=\frac{1}{Z(\b)}\int_0^\infty \d \a_1\,\frac{1}{S_b(\pm \i \a_1/\pi b)}\,\exp\bigg(-(\beta-\t) \frac{\cosh(\a_1 )}{2\pi^2 b^4}\bigg)\nonumber\\&\qquad\int_0^\infty \d \a_2\,\frac{1}{S_b(\pm \i \a_2/\pi b)}\,\exp\bigg(-\t \frac{\cosh(\a_2 )}{2\pi^2 b^4}\bigg)\,\frac{S_b(b\Delta\pm \i \a_1/2\pi b\pm \i \a_2/2\pi b)}{S_b(2b \Delta)}\,.\label{3.22}
\end{align}
It is now convenient to exchange the energy labels $\a_1,\a_2$ for mean energy and energy differences $\a,\g$
\begin{equation}
    \a_1=\a+\g \pi b^2\,,\quad \a_2=\a-\g \pi b^2\,.\label{3.23}
\end{equation}
We claim that the semiclassical limit is obtained by sending $b\to 0$, whilst keeping $\a,\g,\Delta$ finite. This scaling mimics the scaling used to recover the semiclassical limit of the two point function of DSSYK from the exact answer in \cite{Goel:2023svz}. We consider the probe approximation where the perturbation does not backreact heavily on the geometry. This means that the difference of ADM energies of the black holes before and after the infalling particle is small as compared to the black hole mass. For heavy operators $\Delta\sim 1/b^2$ one should consider significant backreaction $\a_1-\a_2\sim 1$.\footnote{The semiclassical two point function \eqref{3.28} indeed does not backreact on the $\a$ saddle as long as $\Delta\ll 1/b^2$.}

Now we enforce this limit on \eqref{3.22}. For the $S_b(x)$ functions we use the following assymptotics \cite{ip2021tensor}\footnote{For $\Delta=n$ one can use the shift relations of $S_b(x)$ to further refine the first equation to a generalization of \eqref{3.1}.}
\begin{equation}
    \lim_{b\to 0}S_b(b\Delta \pm \i \a/\pi b)\sim \sinh(\a)^{2\Delta-1}e^{-\a/b^2}\,,\quad \lim_{b\to 0}S_b(b(\Delta\pm \i \g))\sim \Gamma(\Delta\pm \i \gamma)\,.\label{718888}
\end{equation}
This results in
\begin{equation}
    \lim_{b\to 0}\frac{1}{S_b(\pm \i \a_1/\pi b)}\,\frac{1}{S_b(\pm \i \a_2/\pi b)}\,\frac{S_b(b\Delta\pm \i \a_1/2\pi b\pm \i \a_2/2\pi b)}{S_b(2b \Delta)}\sim \sinh(\a)^{2\Delta+1}e^{\a/b^2}\, \frac{\Gamma(\Delta\pm \i \gamma)}{\Gamma(2\Delta)}\,.
\end{equation}
For the Boltzmann weights we expand to linear order in $\g$
\begin{equation}
    \lim_{b\to 0}\exp\bigg(-(\beta-\t) \frac{\cosh(\a_1 )}{2\pi^2 b^4}-\t \frac{\cosh(\a_2)}{2\pi^2 b^4}\bigg)\sim \exp\bigg( -\beta \frac{\cosh(\a )}{2\pi^2 b^4}+\g \frac{\sinh(\a)}{\pi b^2}\bigg(\tau-\frac{\beta}{2}\bigg)\bigg)\,,
\end{equation}
higher powers in the $\g b^2$ expansion vanish for $b\to 0$. Here we remind the reader that for the semiclassical limit to hold, we are encouraged to consider $\b/\pi b^2\sim 1$ and $\tau/\pi b^2\sim 1$, keeping in mind similar rescaling was necessary to validate the classical approximation for the q-Schwarzian \eqref{rescaledac}. Combining these two elements the two-point function \eqref{3.22} reduces to
\begin{equation}
    \frac{1}{Z(\beta)}\int_0^\infty \d \cosh(\a) \exp\bigg( \frac{\a}{b^2}-\beta \frac{\cosh(\a )}{2\pi^2 b^4}\bigg)\, \sinh(\a)^{2\Delta}\int_{-\infty}^{+\infty} \d \gamma \exp\bigg(\g \frac{\sinh(\a)}{\pi b^2}\bigg(\tau-\frac{\beta}{2}\bigg)\bigg) \frac{\Gamma(\Delta\pm \i \gamma)}{\Gamma(2\Delta)}\,.\label{3.27}
\end{equation}
The $\g$ integral can be computed for generic $\a$ and $\b$ but simplifies on the saddle \eqref{3.5} of the $\a$ integral, which we recognize as $Z(\beta)$ in \eqref{3.4}. Note indeed that for $\g,\Delta\sim 1$ there is no backreaction from the $\g$ integral. Rescaling $\g=\omega \beta/2\pi$, this integral reduces actually to the AdS$_2$ boundary two-point function calculation \cite{Lam:2018pvp,Maldacena:2016upp,Shenker:2014cwa}, with result
\begin{equation}
    \boxed{\average{\mathcal{B}_{b \Delta}\mathcal{B}_{b\Delta}}_{\tau,\,\beta-\tau\,\text{class}}=(-1)^\Delta \frac{\sinh(\a)^{2\Delta}}{\sin(\tau \sinh(\a)/2\pi b^2)^{2\Delta}}=e^{-2 \Delta\varphi(\tau)}\,}\,.\label{3.28}
\end{equation}
This indeed matches the classical solution of $e^{-2\Delta \varphi}$ computed using the q-Schwarzian action \eqref{sch2ptans}.\footnote{Notice that in this equation $0<\tau<\beta$ whereas in \eqref{sch2ptans} $0<\tau<\beta/\pi b^2$, so both equations are indeed identical.} This supports the holographic dictionary \eqref{dic}.

In fact one can convince oneself that the identification \eqref{dic} remains true at the quantum level. To see this, one uses the two-sided q-Liouville version of the q-Schwarzian (see discussion above \eqref{2.6}). Then one can compute expectation values of $e^{2\Delta \phi}$ using canonical quantization and representation theory of quantum groups, by using the slicing illustrated in \eqref{fig:opslice} above. The result matches exactly with \eqref{3.22}. We explain this quantization procedure compactly in section \ref{sect:quant}.

\subsection{Conserved currents and symmetry algebra}\label{sect:2.2}
In the process of solving the equations of motion we already encountered four conserved quantities  $\mu_\b,\mu_\g, \pi_\varphi/2-\beta \pi_\beta$ and $\pi_\varphi/2-\g \pi_\g$. Upon more careful investigation one finds two more conserved charges, bringing the total to six. We are interested in the associated symmetry algebra, which one reads off from the Poisson brackets of the charges. This is more natural to deal with using canonical momenta \eqref{rescale}, and in this section alone we will work with $q=e^{\i\pi b^2}$ \eqref{1.2}. 

One can choose three independent combinations of the conserved currents as
\begin{equation}\label{poissongen}
\begin{aligned}
    &h_L=\frac{1}{2}p_\varphi-\g p_\g\,,
    \\&f_L=-\frac{e^{-2\i \log q\,\g p_\g}-1}{ 2\i\log q\,\g},
    \\& e_L=e^{-2\varphi-i\log q p_\varphi}\,\frac{e^{-2\i \log q\,\b p_\b}-1}{-2\i\log q\,\b}-\g\,e^{\i\log q\, p_\varphi}\,\frac{e^{2\i \log q\,(\g p_\g-p_\varphi)}-1}{2\i\log q}.
\end{aligned}
\end{equation}
The first and second (proportional to $\mu_\g$) we already found, the last one is new. One checks that each of these has vanishing Poisson brackets with $\mathbf{H}(x_A,p_A)$ \eqref{6dimqsch}, or equivalently that their time derivative vanishes. They satisfy a remarkable non-linear algebra
\begin{equation}
    \{h_L,e_L\}=e_L\,,\quad  \{h_L,f_L\}=-f_L\,,\quad  \{e_L,f_L\}=\frac{q^{2\i h_L}-q^{-2\i h_L}}{2\i\log q}\,,\label{algebra}
\end{equation}
which upon quantization becomes the $U_q(\mathfrak{sl}(2,\mathbb{R}))$ quantum algebra \cite{klimyk2012quantum,jaganathan2000introduction}\footnote{As explained in \cite{Blommaert:2023opb} quantization involves scaling $\log q_\text{quant}=\hbar\, \log q_\text{class}$ and for instance $e_L=\i\hbar E$.}
\begin{equation}
    \label{eq:qalg}
    [H,E]=E\,,\quad [H,F]=-F\,,\quad [E,F] = \frac{q^{2H}-q^{-2H}}{q-q^{-1}}\,.
\end{equation}
The three remaining conserved currents are
\begin{equation}\label{q-sch current}
\begin{aligned}
    &h_R=-\frac{1}{2}p_\varphi+\b p_\b\,,
    \\&f_R=e^{-2\varphi-2\i\log q \b p_\b}\,\frac{e^{-2\i \log q\,\g p_\g}-1}{2\i\log q\,\g}-\b\,\frac{e^{2\i \log q\,(p_\varphi-\b p_\b)}-1}{2\i\log q},
    \\& e_R=-\,e^{-\i\log q\, p_\varphi}\,\frac{e^{2\i \log q\,\b p_\b}-1}{2\i\log q\,\b},
\end{aligned}
\end{equation}
where the last one is proportional to $\mu_\beta$ (up to some exponential of $h_R$). These satisfy the same algebra as the generators \eqref{algebra}, and Poisson-commute (i.e. have vanishing Poisson brackets) with them. The q-Schwarzian \eqref{6dimqsch} is essentially the simplest dynamical system with this symmetry, the Hamiltonian $\mathbf{H}(x_A,p_A)$ is (up to trivial rescalings and powers) the unique function that commutes with all these currents. In fact it is the Casimir of $U_q(\mathfrak{sl}(2,\mathbb{R}))$:
\begin{equation}
     \mathbf{H}=e f-\frac{\cos(2\log q h)}{2\log q^2}\,.
\end{equation}
Alternatively phrased, specifying a six-dimensional symmetry algebra $U_{q,L}(\mathfrak{sl}(2,\mathbb{R})) \otimes U_{q,R}(\mathfrak{sl}(2,\mathbb{R}))$ of Poisson brackets, it can be rewritten in the standard canonical form, by Darboux's theorem, up to canonical transformations, fully specifying the dynamical system.

Moreover, after a suitable redefinition (see e.g. equation (4.2) in \cite{Blommaert:2023opb}), the variables $(\varphi,\b,\g)$ can be identified as coordinates on the quantum group SL$_q^+(2,\mathbb{R})$, so our model describes quantum mechanics of a particle on SL$_q^+(2,\mathbb{R})$. 

The one-sided system \eqref{2.6} (the genuine q-Schwarzian) has one constraint
\begin{equation}
    e^{\i\log q h_L}f_L=\frac{-\i}{2\log q}\,,\label{oneconstraint}
\end{equation}
the two-sided system \eqref{2.6} (q-Liouville) has two constraints, namely \eqref{oneconstraint} and
\begin{equation}
    e^{-\i\log q h_R}e_R=\frac{\i}{2\log q}\,.\label{twoconstraint}
\end{equation}
The first allows one to gauge-fix $\g=0$ and the second $\beta=0$ \cite{Blommaert:2023opb}, multiplying both then leads to  \eqref{2.20}: $4\mu_\b\mu_\g=1$.\footnote{Instead of gauge-fixing one could think of fixing $h_L+h_R=0$ which commutes with $4\mu_\g\mu_\b=1$.} Note furthermore that these constraints commute with $\mu_\b\mu_\g e^{-2\varphi}$ such that quantum mechanically the boundary conditions \eqref{2.19} are consistent.

Finally let us remark that in the \emph{target space} of the particle on the group manifold, the quantum group (non-commutative) manifold itself and symmetry generators have geometric interpretations as acting on the trajectory of the quantum particle.\footnote{For $\Uqsu$ (the case $0<q<1$), such a target space picture was discussed in \cite{Berkooz:2022mfk}, this is the space where the chords ``live''.} However, this ``target'' is not where our quantum gravity lives,\footnote{Exceptions to this are JT gravity itself \cite{Yang:2018gdb} and its higher supersymmetric variants \cite{Belaey:2023jtr}, where the bulk geometry and the target space of the boundary dual are both an elliptic (Euclidean signature) or hyperbolic (Lorentzian signature) coset of the SL$(2,\mathbb{R})$ group manifold (or its supersymmetric cousins).} so for our purposes this picture seems of little use. We don't have a crisp understanding of the role of the symmetry transformations in our bulk (dilaton) gravity dual.

\section{Path integral derivation}\label{sect:PIargument}
In this section we will show that the q-Schwarzian theory \eqref{6dimqsch} is exactly equal to sinh silaton gravity \eqref{3.6}. Just like for the ordinary Schwarzian/JT gravity duality, this duality holds only for the leading disk topology in the gravitational theory. Including non-trivial topologies, JT gravity becomes dual to a matrix integral \cite{Saad:2019lba}.\footnote{See also \cite{Blommaert:2022ucs} for a 1d quantum mechanics dual to a factorizing minimal modification of JT gravity.} Presumably a similar story holds for sinh dilaton gravity.\footnote{See  \cite{Collier:2023cyw} for relevant recent work, although they use different boundary conditions and as a result obtain different thermodynamics and amplitudes.}

We will prove this duality through an intermediate theory, a Poisson sigma model with Lorentzian action:
\begin{equation}\label{actionPSM}
    S_\text{PSM}=\int_0^\infty \d u\int\d t\bigg(-(A_u)_B\dot{J_B}+(A_t)_B\bigg(J_B'+\alpha_{BC}(J_A)\, (A_u)_C\bigg)\bigg)-\int\d t\, \mathbf{H}(J_A)\,.
\end{equation}
This is a two-dimensional gauge theory (the sense in which it's a \emph{gauge} theory will become clear). The relevant model for our purposes has a three-dimensional ``target space''.  with gauge connection $(A_\mu)_B = (A_\mu)_0,(A_\mu)_1,(A_\mu)_H$ and $J_A=J_0,J_1,J_H$, where we label (with hindsight) as $A=0,1,H$, and with a time-like boundary at $u=0$. We ordered the coordinates as $\d^2x=\d t\wedge \d u$ and summing over repeated indices is assumed. The $\alpha_{BC}(J_A) \equiv -\alpha_{CB}(J_A)$ are functions of $J_A$\cite{Cattaneo:2001bp}. In principle these functions can be chosen in a model dependent way. For the system we are discussing, they are chosen to be
\begin{equation}
    \alpha_{H 0}=-J_1, \quad \alpha_{H 1}=-J_0, \quad \alpha_{0 1}=\frac{\sinh(2\pi b^2 J_H)}{2\pi b^2}=\frac{1}{2}V_\text{qsch}(J_H)\,.\label{alphaaaa}
\end{equation}
From the first term in \eqref{actionPSM} we observe that the six fields come with (equal-time) fundamental Poisson brackets
\begin{equation}
    \Big\{(A_u)_B(u_1,t_0),J_C(u_2,t_0)\Big\}=\delta_{B C}\,\delta(u_1-u_2)\,.\label{4.15}
\end{equation}
In \textbf{section \ref{sect:topotobdy}} we will show the phase space dynamics of this Poisson sigma model can be reduced to a q-Schwarzian path integral as a boundary description. In \textbf{section \ref{sect:vielbein}} we will show the equivalence between this Poisson sigma model and sinh dilaton gravity. In \textbf{section \ref{sect:com}} we comment on boundary conditions.

\subsection{Poisson sigma model and the q-Schwarzian}\label{sect:topotobdy}
There are two inputs in \eqref{actionPSM} for the specific system we are considering. Firstly, the choice made in \eqref{alphaaaa} is related to the $\Uqsl$ quantum algebra we are studying here. Secondly, we have explicitly added the boundary term $\mathbf{H}(J_A)$, which will play the role of the Hamiltonian of the q-Schwarzian quantum mechanics \eqref{6dimqsch} later on.
We'll show the correspondence between the bulk Poisson sigma model and the boundary q-Schwarzian model in two steps: 
\begin{enumerate}
    \item We will show that the \textbf{dynamics} of \eqref{actionPSM} \textbf{reduces to dynamics on a $\bf{6}$-dimensional phase space} \cite{Cattaneo:2001bp}, which one can think of as living on the spatial boundary. This is a huge reduction from the original naive QFT phase space.
    \item  We show that such \textbf{dynamics is equivalent to the dynamics described by a q-Schwarzian action}. The proof will be done in the classical phase space, but after canonical quantization, also the quantum duality will be proven. 
\end{enumerate}

\textbf{Step one.} In the general case $\a_{BC}(J_A)$ is non-linear, obstructing a simple integrating out of $J_A$, unlike for classical Lie algebras.\footnote{For Lie algebras $\a_{BC}(J_A)=\tensor{f}{_{BC}^A} J_A$, and $A_B$ are the components of the one-form $A$. The bulk action in that case is linear in $J$ and localizes on $A=\d g g^{-1}$ with $g(u,t)$ group-valued, after integrating over $J_A$. The remaining action only depends on boundary values $g(t)$ and $J(t)$ and results in quantum mechanics on the group manifold \cite{Blommaert:2018oro}.} Nevertheless, following \cite{Cattaneo:2001bp} (section 2.1), one can explicitly show that the theory is topological with a $2m$-dimensional classical phase space (in our case $m=3$, the range of the label $A$). The point is that one can integrate out bulk values of $(A_t)_B$, resulting in a \emph{constraint}:
\begin{equation}
    J_B'(u,t_0)=-\sum_C\alpha_{BC}(J_A(u,t_0)) (A_u)_C(u,t_0)\,,\label{4.10}
\end{equation}
where the prime $'$ denotes differentiation w.r.t. the spatial coordinate $u$. This is a constraint on phase space, which one associates with the field content on a Cauchy slice, labeled by the arbitrary time $t_0$. We'll often suppress this label $t_0$ for presentation purposes. We will first determine the physical phase space (prior to quantizing), which means one solves the constraints and mods out by gauge transformations generated by the constraints.  We will explicitly construct solutions to \eqref{4.10} with $2m$ ($m=3$ here) free parameters, which should be gauge-equivalent to any other solution. 

We can construct simple solutions by mapping \eqref{4.10} into the equation of motion of an \emph{auxiliary} classical mechanics problem. Consider a Hamiltonian system with canonical coordinate $X_A(u)$ and $P_A(u)$, evolving along a ``time'' coordinate $u$, with ``Hamiltonian'' $\mathbf{G}(J_A)$, and with phase space action:
\begin{equation}
    \int_0^{+\infty} \d u \bigg( \sum_A P_A X_A'-\mathbf{G}(J_A)\bigg).\label{4.11}
\end{equation}
We want to design this auxiliary system such that the constraint \eqref{4.10} is one of its Hamilton equations. We first go to the non-canonical coordinate system $(X_A,J_A)$, where $J_A(X_A,P_A)$ have ``Poisson brackets''
\begin{equation}
    \bfb J_A,J_B\bfk=\alpha_{AB}(J_C)\,.\label{bulkPB}
\end{equation}
We use a bold symbol $\bfb$ to remind the Poisson brackets here are only for the auxiliary system, which is unrelated to the Poisson bracket of the bulk theory \eqref{4.15}. The latter is relevant for evolution in $t$, whereas our current discussion takes place on one Cauchy slice $t=t_0$.

Now in this non-canonical coordinate system, the Hamilton equations of motion are
\begin{equation}
    J_B'=\sum_C \bfb J_B,J_C \bfk \frac{d\mathbf{G}}{d J_C}, \qquad X_B'=\sum_C \frac{dJ_C}{dP_B}\,\frac{d\mathbf{G}}{dJ_C}\,. \label{Sec.EOM}
\end{equation}
Now suppose we get a solution of the above equation of motion, then let us identify\footnote{Given a solution $J_A(u,t_0)$ the right hand side is an explicit function of $u$.}
\begin{equation}
    (A_u)_C=-\frac{\d \mathbf{G}}{\d J_C}\,.\label{4.13}
\end{equation}
Together with the ``Poisson brackets'' \eqref{bulkPB}, the first equation in \eqref{Sec.EOM} becomes exactly the constraint \eqref{4.10}. But noticing the constraint \eqref{4.10} is actually a boundary value problem, with a boundary at $u=0$. We denote the six boundary condition as $X_A(0)=x_A$ and $J_A(0)=j_A$. This in principle characterizes all possible configurations that satisfy the constraint \eqref{4.10}, for a given \emph{choice} of $G(J_A)$. We'll comment on the meaning of this (freedom of) choice in section \ref{sect:comnew}.
Thus we finished the first step, reduce the dynamics into a $6$-dimensional phase space spanned by $(x_A,j_A)$.

\textbf{Step two.} Next we show that the dynamics is essentially the same as the one of the q-Schwarzian path integral. We start by deriving the Poisson bracket on 6d phase space spanned by $(x_A,j_A)$. For the currents $j_A$, a derivation was given in \cite{Cattaneo:2001bp}. Integrating the constraint \eqref{4.10} along a spatial slice, we get:
\begin{equation}
    j_A(t_0)=J_A(\infty,t_0))+\int_0^\infty\d u_1\, \alpha_{A D}(J_C(u_1,t_0))\, (A_u)_D(u_1,t_0)\,.
\end{equation}
Remember the Poisson bracket of bulk variable is given by \eqref{4.15}. Now we compute the Poisson brackets of the right hand side with $J_B(u_2,t_0)$ (suppressing labels $t_0$ henceforth)
\begin{equation}
\Big\{j_A,J_B(u_2) \Big\} =   \int_0^\infty\d u_1\, \alpha_{A D}(J_C(u_1))\, \Big\{(A_u)_D(u_1),J_B(u_2)\Big\}=\alpha_{A B}(J_C(u_2))\,.
\end{equation}
Considering the special case $J_B(u_2=0)=j_B$ we have
\begin{equation}\label{PBJ}
    \Big\{j_A,j_B\Big\}=\alpha_{AB}(j_C)\,.
\end{equation}
Similarly, we can integrate over \eqref{Sec.EOM} using \eqref{4.13}
\begin{equation}
    x_A=X_A(\infty)+\int_0^\infty \d u_1\,\frac{\d J_D}{\d P_A}\,(A_u)_D(u_1)
\end{equation}
Remember here that $\d J_D/d P_A$ is a function of $X_C(u_1)$ and $J_C(u_1)$. Computing the Poisson bracket of the right hand side with $J_B(u_2)$ using \eqref{4.15} gives
\begin{equation}
    \int_0^\infty \d u_1\,\frac{\d J_D}{\d P_A}\,\Big\{(A_u)_D(u_1),J_B(u_2)\Big\}=\frac{\d J_B}{\d P_A}\big(X_C(u_2),J_C(u_2)\big).
\end{equation}
This again for the special case $J_C(0)=j_C$ and $X_C(0)=x_C$ gives
\begin{equation}
    \Big\{x_A,j_B\Big\}=\frac{\d j_B}{\d p_A}\,.\label{PBXJ}
\end{equation}
This shows that $x_A$ and $p_A$ are just canonical conjugate variables with respect to the original Poisson brackets:
\begin{equation}
    \Big\{x_A,p_B\Big\}=\delta_{A B}\,.
\end{equation}

Noticing furthermore that, having integrated out $(A_t)_B$ in the Poisson sigma model \eqref{actionPSM} the Hamiltonian comes entirely from the explicit boundary term $\mathbf{H}(J_A(0))=\mathbf{H}(j_A)$. We thus have a dynamical system where the Hamiltonian is given by
\begin{equation}
    \mathbf{H}(j_A)=-j_1^2+j_0^2+\frac{\cosh(2\pi b^2 h)}{2\pi^2 b^4}\,,\label{3.15bis}
\end{equation}
and the Poisson brackets are given by \eqref{PBJ} (inserting our specific case \eqref{alphaaaa})
\begin{equation}
    \Big\{h,j_0\Big\}=-j_1\,,\quad \Big\{h,j_1\Big\}=-j_0\,,\quad \Big\{j_0,j_1\Big\}=\frac{\sinh(2 \pi b^2 h)}{2\pi b^2}\,,\label{4.3}
\end{equation}
and \eqref{PBXJ}. We now explain this is nothing but the q-Schwarzian path integral \eqref{6dimqsch} in a non-canonical coordinate. We should identify $x_A=(\varphi,\g,\b)$ and the currents $j_A=(j_0,j_1,h)$, where
$-2j_0=e+f$ and $2j_1=e-f$. Here $(e,f,h)$ are the currents in \eqref{q-sch current}, which means $j_A$ functions of \emph{canonical} coordinates $(x_A,p_A)$. As a result we recover precisely the q-Schwarzian phase space path integral \eqref{6dimqsch}. 
\begin{equation}
\boxed{\begin{aligned}
 &\exp\bigg(\i\int_0^\infty \d u\int\d t\bigg(-(A_u)_B\dot{J_B}+(A_t)_B\bigg(J_B'+\alpha_{BC}(J_A)\, (A_u)_C\bigg)\bigg)-\i \int\d t\, \mathbf{H}(J_A)\bigg)\\&\qquad\qquad\qquad\qquad\qquad\qquad\qquad\qquad\qquad\qquad\quad\Leftrightarrow \quad\exp\bigg( \i \int\d t\bigg(\sum_A p_A \dot{x_A}-\mathbf{H}(j_A)\bigg)\bigg)\,\label{duality}
\end{aligned}}
\end{equation}
The quantization of this system reproduces Liouville gravity amplitudes, as determined in e.g. \cite{Mertens:2020hbs,Fan:2021bwt}, as we explain in section \ref{sect:quant}. The amplitudes are in this language determined in a BF-like fashion as studied earlier in \cite{Mertens:2018fds,Blommaert:2018oue,Blommaert:2018oro,Blommaert:2018iqz,Iliesiu:2019xuh}, in a similar way as in the older 2d Yang-Mills literature \cite{Migdal:1975zg,Witten:1992xu,Cordes:1994fc}. 

This logic extends to generic dilaton potentials. To make the quantum mechanical model explicit, the only difficulty might be to find a system of $m$ canonical pairs $(x_A,p_A)$ that realizes the non-linear algebra \eqref{PBJ}, i.e. finding relations $j_A(x_B,p_B)$.

\subsection{Poisson sigma model and sinh dilaton gravity}\label{sect:vielbein}
Oftentimes quantization of gravitational theories can be accomplished by using the first order (vielbein) formulation, thereby rewriting the gravity model in question as a gauge theory that one knows how to quantize. The most famous examples are AdS$_3$ gravity (see for instance 
\cite{Witten:1988hc,Witten:2007kt,Donnay:2016iyk,Cotler:2020ugk,Collier:2023fwi}) and JT dilaton gravity (see for instance \cite{Fukuyama:1985gg,Isler:1989hq, Chamseddine:1989yz,Jackiw:1992bw,Blommaert:2018oro,Blommaert:2018iqz,Iliesiu:2019xuh,Saad:2019lba}), which in first-order variables are (respectively) 3d $\text{SL}(2,\mathbb{R})\times \text{SL}(2,\mathbb{R})$ Chern-Simons and 2d SL$(2,\mathbb{R})$  BF gauge theories.\footnote{In both cases there are subtleties with this identification having to do with gauging large diffeos in gravity and restrictions that metric invertibility puts on the allowed gauge connections. These subtleties can be largely ignored when one restricts to the simplest bulk topology, which we are doing in this work.} Similarly it is known that general 2d dilaton gravity models classically can be written as Poisson sigma models in the vielbein formulation \cite{Ikeda:1993aj,Ikeda:1993fh,Schaller:1994es,Grumiller:2021cwg,Grumiller:2020elf}. In the case of sinh dilaton gravity, as we showed, this allows for an exact quantization through the reduction to the boundary q-Schwarzian. 

For sinh dilaton gravity \eqref{3.6}, this relation was discussed in part in \cite{Fan:2021bwt,Blommaert:2023opb}. We will be slightly more detailed than those works, especially considering physics at the boundary. The starting point is \eqref{actionPSM}
\begin{align}
    &\exp\bigg(\i\int_0^\infty \d u\int\d t\bigg(-(A_u)_B\dot{J_B}+(A_t)_B\bigg(J_B'+\alpha_{BC}(J_A)\, (A_u)_C\bigg)\bigg)-\i \int\d t\, \mathbf{H}(J_A)\bigg)\nonumber\\
    &\qquad\qquad\qquad =\exp\bigg(\i\int\bigg(J_B\,\d A_B+\frac{1}{2}\alpha_{B C}(J_A)A_B\wedge A_C\bigg)-\i \int\bigg(J_B A_B
    +\d t\,  \mathbf{H}(J_A)\bigg)\bigg)\,,\label{4.23}
\end{align}
where the boundary Hamiltonian is \eqref{3.15bis}.\footnote{We are following the notation of \cite{Fan:2021bwt}, although clearly it makes more sense to think of $J_0$ as spacelike and $J_1$ as timelike.} Now we relabel the variables
\begin{equation}
    \Phi=-J_H\,,\quad \Phi_0=-J_1\,,\quad \Phi_1=-J_0\,,\quad \omega=A_H\,,\quad e^1=A_0\,,\quad e^0=A_1\,,\label{mappp}
\end{equation}
such that the Poisson sigma model's bulk action part (on the second line of \eqref{4.23}) becomes 
\begin{align}
    &\exp\bigg(\i \int\bigg( J_0 \d A_0+J_1\d A_1+J_H\d A_H-J_1\, A_H\wedge A_0-J_0\,A_H\wedge A_1+\frac{\text{sinh}(2\pi b^2 J_H)}{2\pi b^2}A^0\wedge A^1\bigg)\bigg)\nonumber\\
    &\quad=\exp\bigg( \i \int\bigg( -\Phi \d \omega +\frac{\text{sinh}(2\pi b^2 \Phi)}{2\pi b^2} e^0\wedge e^1-\Phi_0(\d e^0-\omega\wedge e^1)-\Phi_1(\d e^1-\omega\wedge e^0)\bigg)\bigg)\,.\label{4.26}
\end{align}
Furthermore the boundary term on the second line of \eqref{4.23} evaluates to
\begin{equation}
    \exp\bigg(\i \int\bigg( \Phi \omega +\Phi_0 e^0+\Phi_1 e^1- \d t\, \mathbf{H}(\Phi_0,\Phi_1,\Phi)\bigg)\bigg)\,.\label{4.27}
\end{equation}
This can be rewritten as a dilaton gravity model by combining the zweibeins into a 2d metric\footnote{Conventions for Lorentzian signature include
\begin{equation}
    \eta_{00}=-1\,,\quad \eta_{11}=1\,,\quad \varepsilon_{t u}=\varepsilon_{0 1}=1\,,\quad \omega^{a b}=\varepsilon^{ a b} \omega\,.
\end{equation}
The Levi-Cevita symbol appears for instance in the relation
\begin{equation}
    \d^2x=\d t\wedge \d u=\frac{1}{2}\varepsilon_{\mu\nu}\d x^\mu\wedge \d x^\nu=\frac{1}{2 e}\varepsilon_{a b}e^a\wedge e^b=\frac{1}{e}e^0\wedge e^1\,.
\end{equation}
In these coordinates the Levi-Cevita tensor and symbol are identical because $\sqrt{-\eta}=1$.}
\begin{equation}
    g_{\mu\nu}=\eta_{a b}e^a_\mu e^b_\nu\,.
\end{equation}
The curvature two form in two dimensions is
\begin{equation}
    R^{a b}=\d \omega^{a b}+\tensor{w}{^a_b}\wedge w^{c b}=\d \omega^{a b}=\varepsilon^{ a b} \ d\omega
\end{equation}
Furthermore in two dimensions this tensor is fully determined by the Ricci scalar \cite{Saad:2019lba}
\begin{equation}
    R^{a b}=\frac{1}{2}R\,e^a\wedge e^b =-\frac{1}{2}\varepsilon^{a b} \d e^0\wedge \d e^1=-\frac{1}{2}\varepsilon^{ a b} R \sqrt{-g}\d^2 x\,.
\end{equation}
Combining these expressions results in
\begin{equation}
    \frac{1}{2}\sqrt{-g} R \d^2 x=-\d\omega\,.
\end{equation}
Furthermore the torsion constraints \cite{Donnay:2016iyk} reduce to
\begin{equation}
    T^a=\d e^a+\tensor{\omega}{^a_b} \wedge e^ b=\d e^a+\varepsilon^{a b} \omega \wedge e_b=0\,,
\end{equation}
Combining these elements we find that the bulk part of the Poisson sigma model action \eqref{4.26}  reduces to the bulk action of sinh dilaton gravity \eqref{3.6}
\begin{align}
    &\exp\bigg( \i \int\bigg( -\Phi \d \omega +\frac{\text{sinh}(2\pi b^2 \Phi)}{2\pi b^2} e^0\wedge e^1-\Phi_0(\d e^0-\omega\wedge e^1)-\Phi_1(\d e^1-\omega\wedge e^0)\bigg)\bigg)\nonumber\\
    &\qquad\qquad\qquad\qquad\qquad\qquad=\exp\bigg(\i\int \d^2 x \sqrt{-g}\,\bigg(\frac{1}{2}\Phi R + \frac{\text{sinh}(2\pi b^2 \Phi)}{2\pi b^2}-\Phi_0 T^0-\Phi_1 T^1\bigg)\bigg)\label{3.29}
\end{align}
Usually one would not write the last two terms, since $T^a=0$. We keep it explicit for future discussions. 
Concerning the boundary term \eqref{4.27} one finds that the first term is the usual Gibbons-Hawking-York curvature term
\begin{equation}
    \exp\bigg(\i \int\bigg( \Phi \omega +\Phi_0 e^0+\Phi_1 e^1- \d t\, \mathbf{H}(\Phi_0,\Phi_1,\Phi)\bigg)\bigg)=\exp\bigg(\i \int \d t\,\sqrt{-h}\,\Phi K+\text{counterterms}\bigg)\,.\label{4.36}
\end{equation}

We now comment on the additional boundary terms in \eqref{4.36} and clarify why they can be interpreted as counterterms. For this we will return to Euclidean signature, and consider the classical solutions \eqref{3.7} of the bulk theory \eqref{3.29}, which we repeat here for reader comfort:
\begin{equation}
    \d s^2=F(r)\d\t^2+\frac{\d r^2}{F(r)}\,,\qquad \Phi=r\,,\qquad F(r)=\frac{\cosh(2\pi b^2 r)}{2\pi^2 b^4}-\frac{\cosh(2\pi b^2 \Phi_h)}{2\pi^2 b^4}\,.\label{4.37}
\end{equation}
In this gauge, one can take $\Phi_1=0$ and solve the $\omega$ EOM in \eqref{4.26} for
\begin{equation}
    \Phi_0=\frac{1}{e^1_r}=F(r)^{1/2}\,.\label{4.38}
\end{equation}
Plugging the explicit solution for $F(r)$ into the boundary Hamiltonian, one recovers the (finite) on-shell energy of the q-Schwarzian system \eqref{enenen}
\begin{equation}
    H(\Phi_0,\Phi_1,\Phi)=-F(r)+\frac{\cosh(2\pi b^2 r)}{2\pi^2 b^4}=\frac{\cosh(2\pi b^2 \Phi_h)}{2\pi^2 b^4}\,.\label{4.39}
\end{equation}
This was expected, as we designed $H(\Phi_0,\Phi_1,\Phi)$ to become the q-Schwarzian Hamiltonian in this duality, nevertheless it is reassuring. To reconcile this with previous statements about counterterms, we should include the on-shell contribution from the boundary term $\Phi_0 e^0$ in \eqref{4.36}. We see that this cancels the $F(r)$ term in \eqref{4.39}. On-shell one then indeed identifies the additional terms in \eqref{4.36} as a counterterm
\begin{equation}
    \exp\bigg(\int\bigg( \i\Phi \omega +\i\Phi_0 e^0+\i\Phi_1 e^1- \d \t\, \mathbf{H}(\Phi_0,\Phi_1,\Phi)\bigg)\bigg)=\exp\bigg(\int \d \t\,\sqrt{h}\,\bigg(\Phi K-\sqrt{\frac{\cosh(2\pi b^2 \Phi)}{2\pi^2 b^4}}\bigg)\bigg)\,.
\end{equation}
In fact, looking back at our evaluation of the sinh dilaton gravity on-shell action \eqref{3.6}, we notice that this is precisely the counterterm required to make the ADM energy \eqref{3.15} finite and equal to \eqref{4.39}.
\begin{align}
     &\int \dpi \Phi \dpi g\,\exp\bigg( \frac{1}{2}\int \d x \sqrt{g}\bigg(\Phi R + \frac{\sinh(2\pi b^2\P)}{\pi b^2}\bigg)+\int \d \t\,\sqrt{h}\,\bigg(\Phi K-\sqrt{\frac{\cosh(2\pi b^2 \Phi)}{2\pi^2 b^4}}\bigg)\bigg)\nonumber\\&\qquad\qquad\qquad\qquad\qquad\qquad\qquad\qquad\qquad\qquad \overset{\text{class}}{=}\int_0^\infty \d \Phi_h \exp\bigg(2\pi \Phi_h-\b \frac{\cosh(2\pi b^2 \Phi_h)}{2\pi^2 b^4}\bigg)\,.
\end{align}
It is reassuring that our derivation reproduces this counterterm, which strictly speaking was necessary to obtain a match between the classical bulk-and boundary theories in section \ref{sect:thermo}.

We would like to emphasize that in principle this derivation can immediately be used to write down an exact 1d quantum mechanical boundary dual of dilaton models with generic potentials $V(\Phi)$, by substituting in \eqref{alphaaaa} and putting more generally the boundary Hamiltonian \cite{Klosch:1995fi}
\begin{equation}
    \mathbf{H}(J_A)=-J_1^2+J_0^2+\int^{H}\d \Phi\,V(\Phi)\,.
\end{equation}
Then one could attempt to find WdW wavefunctions via the dual quantum mechanics, and reproduce the spectral density of \cite{Maxfield:2020ale,Witten:2020wvy}.

\subsection{Boundary conditions}\label{sect:com}
When quantizing the q-Schwarzian, we reproduce Liouville gravity (for $\abs{q}=1$) \cite{Fan:2021bwt} respectively DSSYK (for $q<1$) \cite{Blommaert:2023opb}, but only after we diagonalize a combination of the currents \eqref{q-sch current}\footnote{See equation (3.34) in \cite{Blommaert:2023opb} and (2.27) in \cite{Fan:2021bwt}.}
\begin{equation}
    -e^{\pi b^2 h}f=\frac{1}{2\pi b^2}\,.
\end{equation}
At first glance this is a peculiar looking constraint, but after using the gravitational dictionary it makes perfect sense. In terms of \eqref{mappp}, eliminating $\Phi_0$ via \eqref{4.38}, one can write this as a boundary condition
\begin{equation}
    \boxed{\sqrt{h}\,e^{-\pi b^2 \Phi}=\frac{1}{2\pi b^2}}\,. \label{constraint}
\end{equation}
One can recognize this as precisely the standard fixed length boundary conditions in 2d Liouville CFT \eqref{3.21} \cite{Fateev:2000ik}, discussed in this context for instance in \cite{Mertens:2020hbs}
\begin{equation}
    \ell=\int_0^\beta\d \tau e^{b\varphi}=\frac{\beta}{2\pi b^2}\,\text{ fixed,}\quad e^{b\varphi}=\sqrt{h}\,e^{-\pi b^2 \Phi}\,.
\end{equation}
The sinh dilaton gravity interpretation is that this is specifying fall-off conditions for the metric near the asymptotic boundary, the analogue of Brown-Henneaux boundary conditions \cite{brown1986central}
\begin{equation}
    F(r)=\frac{e^{2\pi b^2 r}}{2\pi^2 b^4}+ \text{ finite,}\qquad r\to+\infty\,.
\end{equation}
This is indeed the asymptotic behavior or our classical solutions \eqref{3.7}.\footnote{Choosing smaller constants on the right of \eqref{constraint} would appear to place the boundary classically at a finite cutoff radius $r_\text{bdy}$.}

\subsection{Further comments}\label{sect:comnew}
We finish this section with two comments about our holographic construction
\begin{enumerate}
    \item For Lie groups
    \begin{equation}
        A_C=\sum_A \d X_A\,\frac{\d P_A}{\d J_C}\,,\label{4.22}
    \end{equation}
    reduces to $A=\d g g^{-1}$. Indeed in that case $P_A$ is linear in $J_C$, so the right hand side only depends on the coordinates $X_A$. For BF, this is the most general solution of the constraint. With this BF analogy it is tempting to speculate that the full Poisson sigma model might also localize to \eqref{4.22} for generic profiles $X_A(u,t), J_A(u,t)$. We now argue that this must indeed happen. The key point is that evaluating the Poisson sigma model action \eqref{actionPSM} on these configurations \eqref{4.22} actually reproduces the q-Schwarzian action \eqref{6dimqsch}, independent of the bulk profiles of $X_A(u,t), J_A(u,t)$. Moreover we just argued that special cases of \eqref{4.22} are included in the locus of the Poisson sigma model. As all configurations \eqref{4.22} have equal action, they must all be included in the locus. We did not rule out that besides \eqref{4.22} there are other equal action configurations, but for BF there are none, supporting localization to \eqref{4.22}.
    \item  Given the choice of $\textbf{G}(J_A)$, the system is ``constructively'' holographic. By this we mean that for a boundary configuration $x_A(t),j_A(t)$ we can reconstruct the bulk configuration $A_B(u,t),J_C(u,t)$. In this sense the system is similar to the way the ordinary Schwarzian and JT are related, or how 3d Chern-Simons and 2d WZW are related.
    
    With an explicit map \eqref{mappp} between theories it's worth revisiting this somewhat abstract solution \eqref{Sec.EOM} and \eqref{4.13} to the constraints of the gauge theory. The choice of generator $\mathbf{G}(J_A)$ does not affect the action of the solution \eqref{6dimqsch}, so it is a gauge choice. This is related with diffeo invariance in gravity, part of this choice of picking a Hamiltonian to describe radial evolution is the choice of what the radial coordinate $u$ actually is. Given such a choice the components of $A_r$ are determined in terms of the classical solutions of the currents by equation \eqref{4.13}
    \begin{equation}
        e^1_r=\frac{\d \mathbf{G}}{\d\Phi_1}\,,\quad e^0_r=\frac{\d \mathbf{G}}{\d\Phi_0}\,,\quad \omega_r=\frac{\d \mathbf{G}}{\d\Phi}\,.\label{4.46}
    \end{equation}
    As the components of $A_t$ were integrated out, they play no role in this discussion. We partially fix the gauge by choosing $\mathbf{G}$ to be independent of $\Phi$, such that $\omega_r=0$. Now the constraint equations \eqref{4.10} read
    \begin{equation}
        \Phi'=\Phi_0\, e^1_r+\Phi_1\,e^0_r\,,\quad \Phi_0'=\frac{\sinh(2\pi b^2\P)}{2\pi b^2}\, e^1_r\,,\quad \Phi_1'=\frac{\sinh(2\pi b^2\P)}{2\pi b^2}\,e^0_r\,.\label{4.47}
    \end{equation}
    We now choose $\mathbf{G}=\Phi_1/\Phi_0$\,, which we'll see corresponds to the usual gauge-choice \cite{Witten:2020ert} of dilaton gravity which we've used in \eqref{4.37} and \eqref{4.38}. The last equation admits the trivial solution $\Phi_1=0$ and the first equation reduces to $\Phi'=1$, which admits $\Phi=r$. Furthermore \eqref{4.46} implies $e^0_r=0$ and $e^1_r=1/\Phi_0$. Defining $e^1_r=\sqrt{F}$ the second equation of \eqref{4.47} finally reduces to 
    \begin{equation}
        F'=\frac{\text{sinh}(2\pi b^2 r)}{\pi b^2}.\label{4.48}
    \end{equation}
    This indeed is the radial evolution equation that builds the bulk metric, we recover this as solution of the auxiliary classical mechanics system \eqref{4.11}, which we claim to be playing the role of ``bulk reconstruction'' in our model. We remark that in the sigma model \eqref{4.48} stems from a constraint, begging the question whether sinh dilaton gravity in fact localizes on spacetimes with this metric. It is worth noting that on this classical solution (where $\Phi_1=0$) the radial ``Hamiltonian'' vanishes. This also holds for $\mathbf{G}=\Phi_1$ which corresponds with the gauge $e^1_r=1$, where one finds\footnote{This matches indeed the equation of motion for dilaton gravity in this gauge choice \cite{Stanford:2020wkf}.}
    \begin{equation}
        \Phi''+\frac{\text{sinh}(2\pi b^2 \Phi)}{2\pi b^2}=0\,.
    \end{equation}
    This on-shell vanishing of the radial generator is a consequence of the underlying diff invariance (part of which is $u$ reparameterization ivariance), analogous to the WdW contraint for $t$ evolution.
\end{enumerate}
\section{Concluding remarks}\label{sect:conc}
We conclude this work with several remarks. First, we will briefly touch on a similar duality between DSSYK, the q-Schwarzian ($0<q<1$) and sine dilaton gravity, the details of which will be presented elsewhere. Then, we briefly discuss aspects of quantizing the q-Schwarzian \eqref{6dimqsch} and how this reproduces the quantum amplitudes of Liouville gravity. Finally we highlight a puzzle concerning the boundary two-point function.
\subsection{Closer towards a gravitational dual of DSSYK}\label{sect:dssyk}
The solution of the relevant q-Schwarzian theory is
\begin{equation}
e^{-2\varphi}=\frac{\sin(\theta)^2}{\sin^2(\sin (\theta)\t/2+d)}\,.\label{dssyk5.1}
\end{equation}
Importantly DSSYK comes with the constraint $\varphi> 0$ (in the quantum theory this is saying there is no such think as negative chord number $n$), which one can view as part of the definition of the q-Schwarzian path integral for $0<q<1$. Said differently, the initial and final state is $\varphi=0$. This fixes $d=\theta$. This has quite dramatic consequences as the period (inverse temperature) of the solution is smaller than the ``tomperature'' governing the decay of physical correlators, which remains finite even for $\beta=0$ \cite{Lin:2022nss,Narovlansky:2023lfz}. This constraint (which reduces the period) is crucial to match the on-shell action of the q-Schwarzian with DSSYK thermodynamics.

The derivation of section \ref{sect:PIargument} holds for any dilaton  potential, this includes the duality of DSSYK with a sine dilaton gravity \cite{Blommaert:2023opb,Goel:2022pcu}. This raises the question what the bulk origin of the difference between tomperature and temperature is. For smooth horizons the temperature (period) equals the tomperature (Hawking temperature), suggesting some conical deficit \cite{Mertens:2019tcm} on the horizon. An ``energy-dependent'' defect (this should be imposed as a differential equation since $\g, \theta$ are canonical conjugates)
\begin{equation}    \mathcal{V}_\gamma=\int\d x \sqrt{g}\,e^{-(2\pi-\gamma)\Phi}\nonumber\,,\quad \gamma=2\pi-4\theta\, ,
\end{equation}
inserted into sine dilaton gravity indeed (classically) reproduces DSSYK semiclassics\footnote{We thank Herman Verlinde and Vladimir Narovlansky for discussions on this point.}
\begin{equation}
    \begin{tikzpicture}[baseline={([yshift=-.5ex]current bounding box.center)}, scale=0.6]
 \pgftext{\includegraphics[scale=1]{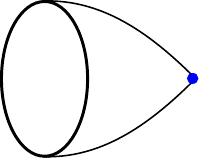}} at (0,0);
    \draw (2.2,0) node {\color{blue}$\mathcal{V}_\gamma$};
    \draw (-2.1,0.5) node {$\beta$};
  \end{tikzpicture}= \int \d \Phi_h\,\exp\bigg(\gamma\Phi_h+\beta \frac{\cos(2\abs{\log q}\Phi_h)}{2\log q^2}\bigg)\,,\quad 2\abs{\log q}\Phi_h=\theta\,.\nonumber
\end{equation}
The near-horizon geometry of this spacetime is similar to the fake disk \cite{Lin:2023trc}. For the boundary two-point function there is a defect both sides of the particle trajectory. Many questions remain, and will not be answered in this appetizer.

\subsection{Matching with Liouville gravity - Quantization}\label{sect:quant}
We briefly discuss the quantization of the q-Schwarzian theory \eqref{6dimqsch}. Following the logic of \cite{Blommaert:2023opb,Fan:2021bwt,Mertens:2022aou} one can reduce this problem to computing the representation matrices of the (modular double \cite{Faddeev:1999fe,Ponsot:1999uf,Ponsot:2000mt,Kharchev:2001rs,Bytsko:2002br,Bytsko:2006ut}) quantum group SL$_q^+(2,\mathbb{R})$. Here we'll employ a more direct presentation. We will consider the two-sided system, where we have to impose two (Brown-Henneaux type) constraints \eqref{twoconstraint}, reducing the system to a 2d phase space path integral with conjugate pair $(\varphi,p_\varphi)$ and with Hamiltonian (from \eqref{6dimqsch})
\begin{equation}
    4\pi^2 b^4 \mathbf{H}(\varphi,p_\varphi)=e^{-\pi b^2 p_\varphi}+(1-e^{-2\varphi})\,e^{\pi b^2 p_\varphi}\,,
\end{equation}
where we used \eqref{2.20} $4\mu_\b\mu_\g=1$ (which follows from the Brown-Henneaux boundary conditions). Using standard canonical quantization
\begin{equation}
    p_\varphi=-\i\hbar \frac{\d}{\d \varphi}\,,
\end{equation}
(and remembering that $\hbar b^2_\text{class}=b^2_\text{quant}$ \cite{Blommaert:2023opb}), this results in the Schr\"odinger equation 
\begin{equation}
    \psi_\alpha(\varphi+\i\pi b^2)+(1-e^{-2\varphi})\,\psi_\alpha(\varphi-\i\pi b^2)=2\cosh(\a)\, \psi_\alpha(\varphi)\,,\label{5.3}
\end{equation}
where we parameterized energy eigenvalues as for instance in \eqref{enenen} and \eqref{onshellqsch}
\begin{equation}
    E(\alpha)=\frac{\cosh(\a)}{2\pi^2b^4}\,.
\end{equation}
This is indeed the same equation that (Whittaker) representation matrix elements of SL$_q^+(2,\mathbb{R})$ satisfy, which is equation (2.46) in \cite{Fan:2021bwt}, with $\varphi=2 b x$.\footnote{In fact, to uniquely find a solution $\psi_\a(\varphi)$ one must realize that they satisfy a second ``dual'' Schr\"odinger equation with $b\to 1/b$. Such duality is not obvious from the classical q-Schwarzian action, but that is no surprise, this symmetry is also notoriously mysterious from the point of view of the classical Liouville (gravity) action.} The solution is given in (2.45) of \cite{Fan:2021bwt}. It suffices here to say that is is intuitively obvious that it can be constructed using the same double-sine function that appears in \eqref{3.1}, and which essentially diagonalizes the type of shifts in $\varphi$ that appear in \eqref{5.3}
\begin{equation}
    S_b(\i \varphi/\pi b+b)=2\i \sinh(\varphi) S_b(\i \varphi/\pi b)\,.
\end{equation}
For our purposes we require the following properties of the wavefunctions \cite{Fan:2021bwt} (up to overall constants)
\begin{equation}
   \int_{-\infty}^{+\infty}\d\varphi\,\psi_{\a_1}(\varphi)\,\psi_{\a_2}(\varphi)^* =\delta(\a_1-\a_2)\, S_b(\pm \i \a_1/\pi b)\,,
\end{equation}
which determines the density of states $\rho(\a)=1/S_b(\pm \i \a/\pi b)$, and the matrix element of $e^{-2\Delta\varphi}$ \eqref{dic}
\begin{equation}
    \int_{-\infty}^{+\infty}\d\varphi\,\psi_{\a_1}(\varphi)\,\psi_{\a_2}(\varphi)^*e^{-2\Delta\varphi}=\frac{S_b(b\Delta\pm \i \a_1/2\pi b\pm \i \a_2/2\pi b)}{S_b(2b \Delta)}\,.
\end{equation}
Finally we need the somewhat nontrivial property that for $\varphi=-\infty$ the wavefunctions are independent of energy $\alpha$.\footnote{This can be deduced directly from the Schrodinger equation \eqref{5.3}, which for $\varphi\to -\infty$ reduces to
\begin{equation}
    e^\varphi\,\psi_\alpha(\varphi+\i\pi b^2)+e^{-\varphi}\,\psi_\alpha(\varphi-\i\pi b^2)=0\,,
\end{equation}
resulting in a pure (rapidly changing) phase. The same reasoning for JT gives $I_0(e^{-\varphi})$ which indeed captures the $\varphi\to -\infty$ assymptotics of $K_{\i\a}(e^{-\varphi})$.
} This is important. As mentioned around \eqref{2.19}, this two-sided q-Schwarzian comes with the specification of an initial (and final) state $\varphi=-\infty$. The partition function is then computed as
\begin{equation}
    \bra{-\infty}e^{-\beta\, \mathbf{H}(\varphi,p_\varphi)}\ket{-\infty}=\int_0^\infty \d\alpha\,\rho(\alpha)\,e^{-\beta E(\alpha)}\,,
\end{equation}
where we've inserted a completeness relation and used the fact that $\psi_\a(-\infty)$ is independent of $\alpha$ (and in fact a phase). This matches the Liouville gravity disk \eqref{3.0}. The Liouville gravity two-point function \eqref{3.22} is similarly reproduced by computing
\begin{equation}
    \bra{-\infty}e^{-(\beta-\tau)\, \mathbf{H}(\varphi,p_\varphi)}\,e^{-2\Delta\varphi}\,e^{-\tau\, \mathbf{H}(\varphi,p_\varphi)}\ket{-\infty}\,.\label{4.11bis}
\end{equation}
The single-sided wavefunctions have also been computed explicitly in \cite{Mertens:2022aou}. These have generic eigenvalue on the inner boundary, which is now a parameter labeling edge  states; see e.g. \cite{Blommaert:2018iqz,Mertens:2022ujr,Wong:2022eiu} for discussions on such edge states in JT and 3d gravity contexts. These wavefunctions are building blocks that give access to OTOC 4-point functions (among other things).  Alternatively, a boundary 4-point function could be computed from the (worldsheet) Liouville gravity perspective, using its 3d TQFT description \cite{Collier:2023cyw}.\footnote{One would have to extend the results of \cite{Collier:2023cyw} to incorporate boundary correlators, possibly using e.g. \cite{Felder:1999mq}.} It would be interesting to match (or compare) such calculations.

\subsection{Two-point function puzzle}\label{sect:5.3}
Finally, we will briefly return to the semiclassical two-point function, which we computed in \eqref{sch2ptans} and \eqref{3.28}. What we \emph{do} know (see the previous subsection and \cite{Fan:2021bwt,Blommaert:2023opb}) is that it computes the expectation value of
\begin{equation}
    \mathcal{W}(\varphi)=R_{\Delta\,\i\i}(e^{2\varphi H})=e^{-2\Delta\varphi}\,,\label{5.1}
\end{equation}
where $H$ is the hyperbolic generator of the quantum algebra \eqref{eq:qalg}.
Usually in BF one considers matrix elements of line integrals of $A$, if the line is placed along a geodesic then
\begin{equation}
    \mathcal{P}\exp\bigg(\int \d u\, A_u\bigg)=\mathcal{P}\exp\bigg(\int\d s\,H\bigg)=e^{\ell H}\,,
\end{equation}
suggesting $2\varphi\overset{?}{=}\ell$. The length $\ell$ of the Einstein-Rosen bridge at $t=0$ in our geometry \eqref{3.7} is finite.
For large black holes $\sinh(\a)\gg 1$, it matches with \eqref{sch2ptans} $2\varphi= -2\log \sinh(\a)$, however one would not expect to require $\sinh(\a)\gg 1$ to match (the saddle is trustworthy also for $\sinh(\a)\sim 1$). Moreover, the fact that $\ell$ remains finite, combined with the fact that we have an initial condition $\varphi=-\infty$, seems to contradict a simple identification between $2\varphi$ and $\ell$.

In fact the dilaton gravity interpretation of \eqref{3.28} is very mysterious. Supposedly we are computing a boundary two point function in sinh dilaton gravity \eqref{3.6}. But the classical answer resembles a pure AdS$_2$ boundary two point function (in terms of $\beta$). How is this possible? Here are two loose thoughts
\begin{enumerate}
    \item For the DSSYK case the answer \eqref{dssyk5.1} was given a geometric interpretation using a ``fake disk'' \cite{Lin:2023trc}. which (as just explained) we think is the actual near horizon geometry of the gravitational dual. The operators are inserted on the AdS$_2$ boundary of that disk. Is there a sense in which correlators with low energy in black holes with a near-AdS$_2$ near-horizon region are accurately approximated by boundary correlators in AdS$_2$ holography?\footnote{Another way to ask this, is if an analysis such as that of \cite{Iliesiu:2020qvm,Almheiri:2016fws,Nayak:2018qej,Moitra:2018jqs} extends to matter probes on the holographic boundary?}
    \item Recently \cite{Narovlansky:2023lfz} a different type of correlator was introduced in the DSSYK context, in the Liouville language corresponding (roughly) to having FZZT boundary conditions on both Liouville fields. This changes the answer significantly, and it that case at least a match with a gravity calculation was found (albeit a 3d gravity one). It could be that such combinations are more natural to study from a gravitational point of view. After all, our choice to study $e^{-2\Delta \varphi}$ (although motivated by JT gravity, and very natural from the group theoretic point of view) was somewhat arbitrary.
\end{enumerate}
One final comment is that it \emph{is} clear that $\varphi$ is the length operator in the target space of the q-Schwarzian (the quantum group manifold SL$_q^+(2,\mathbb{R})$), which one can think of as non-commutative AdS$_2$ \cite{Berkooz:2022mfk}. From that perspective it \emph{is} obvious why $\varphi$ looks like the AdS$_2$ boundary two-point function. Unfortunately, target space is not obviously related with the gravitational bulk, at least from our perspective.

Several groups are interpreting the chord diagrams of DSSYK as a definition of some UV complete quantum gravity in (what we call) target space \cite{Berkooz:2022mfk,Lin:2023trc}, which is reasonable, but it is not in an obvious way the same as the (dilaton) gravity duality we've investigated. Nevertheless it can not be ruled out that the two deceptions are in fact identical. Perhaps integrating out the dilaton in our model leads to a non-commutative bulk?

\section*{Acknowledgments}
We thank Alejandro Cabo-Bizet, Damian Galante, Jorrit Kruthoff, Ohad Mamroud, Vladimir Narovlansky and Herman Verlinde for useful discussions. AB was supported by the ERC-COG Grant NP-QFT No. 864583 and INFN Iniziativa Specifica GAST, and thanks Princeton, Berkeley and CERN for hospitality during parts of this project. TM acknowledges financial support from the European Research Council (grant BHHQG-101040024). Funded by the European Union. Views and opinions expressed are however those of the author(s) only and do not necessarily reflect those of the European Union or the European Research Council. Neither the European Union nor the granting authority can be held responsible for them.
\appendix

\bibliographystyle{ourbst}
\bibliography{Refs}

\providecommand{\href}[2]{#2}\begingroup\raggedright\begin{thebibliography}{10}

\bibitem{jackiw1985lower}
R.~Jackiw, ``Lower dimensional gravity,'' {\em Nuclear Physics B} {\bfseries
  252} (1985) 343--356.

\bibitem{teitelboim1983gravitation}
C.~Teitelboim, ``Gravitation and hamiltonian structure in two spacetime
  dimensions,'' {\em Physics Letters B} {\bfseries 126} no.~1-2, (1983) 41--45.

\bibitem{Maldacena:2016upp}
J.~Maldacena, D.~Stanford, and Z.~Yang, ``{Conformal symmetry and its breaking
  in two dimensional Nearly Anti-de-Sitter space},''
  \href{http://dx.doi.org/10.1093/ptep/ptw124}{{\em PTEP} {\bfseries 2016}
  no.~12, (2016) 12C104}, \href{http://arxiv.org/abs/1606.01857}{{\ttfamily
  arXiv:1606.01857 [hep-th]}}.

\bibitem{Engelsoy:2016xyb}
J.~Engels\"oy, T.~G. Mertens, and H.~Verlinde, ``{An investigation of AdS$_{2}$
  backreaction and holography},''
  \href{http://dx.doi.org/10.1007/JHEP07(2016)139}{{\em JHEP} {\bfseries 07}
  (2016) 139}, \href{http://arxiv.org/abs/1606.03438}{{\ttfamily
  arXiv:1606.03438 [hep-th]}}.

\bibitem{Jensen:2016pah}
K.~Jensen, ``{Chaos in AdS$_2$ Holography},''
  \href{http://dx.doi.org/10.1103/PhysRevLett.117.111601}{{\em Phys. Rev.
  Lett.} {\bfseries 117} no.~11, (2016) 111601},
  \href{http://arxiv.org/abs/1605.06098}{{\ttfamily arXiv:1605.06098
  [hep-th]}}.

\bibitem{Blommaert:2023opb}
A.~Blommaert, T.~G. Mertens, and S.~Yao, ``{Dynamical actions and
  q-representation theory for double-scaled SYK},''
  \href{http://arxiv.org/abs/2306.00941}{{\ttfamily arXiv:2306.00941
  [hep-th]}}.

\bibitem{Cotler:2016fpe}
J.~S. Cotler, G.~Gur-Ari, M.~Hanada, J.~Polchinski, P.~Saad, S.~H. Shenker,
  D.~Stanford, A.~Streicher, and M.~Tezuka, ``{Black Holes and Random
  Matrices},'' \href{http://dx.doi.org/10.1007/JHEP05(2017)118}{{\em JHEP}
  {\bfseries 05} (2017) 118}, \href{http://arxiv.org/abs/1611.04650}{{\ttfamily
  arXiv:1611.04650 [hep-th]}}. [Erratum: JHEP 09, 002 (2018)].

\bibitem{berkooz2018chord}
M.~Berkooz, P.~Narayan, and J.~Simon, ``Chord diagrams, exact correlators in
  spin glasses and black hole bulk reconstruction,'' {\em Journal of High
  Energy Physics} {\bfseries 2018} no.~8, (2018) 1--39.

\bibitem{berkooz2019towards}
M.~Berkooz, M.~Isachenkov, V.~Narovlansky, and G.~Torrents, ``Towards a full
  solution of the large n double-scaled syk model,'' {\em Journal of High
  Energy Physics} {\bfseries 2019} no.~3, (2019) 1--72.

\bibitem{Goel:2022pcu}
A.~Goel, {\em {Investigations of Holographic Duality in Two Dimensions}}.
\newblock PhD thesis, Princeton U., 2022.

\bibitem{Mertens:2020hbs}
T.~G. Mertens and G.~J. Turiaci, ``{Liouville quantum gravity -- holography, JT
  and matrices},'' \href{http://dx.doi.org/10.1007/JHEP01(2021)073}{{\em JHEP}
  {\bfseries 01} (2021) 073}, \href{http://arxiv.org/abs/2006.07072}{{\ttfamily
  arXiv:2006.07072 [hep-th]}}.

\bibitem{Fan:2021bwt}
Y.~Fan and T.~G. Mertens, ``{From quantum groups to Liouville and dilaton
  quantum gravity},'' \href{http://dx.doi.org/10.1007/JHEP05(2022)092}{{\em
  JHEP} {\bfseries 05} (2022) 092},
  \href{http://arxiv.org/abs/2109.07770}{{\ttfamily arXiv:2109.07770
  [hep-th]}}.

\bibitem{Kyono:2017pxs}
H.~Kyono, S.~Okumura, and K.~Yoshida, ``{Comments on 2D dilaton gravity system
  with a hyperbolic dilaton potential},''
  \href{http://dx.doi.org/10.1016/j.nuclphysb.2017.07.013}{{\em Nucl. Phys. B}
  {\bfseries 923} (2017) 126--143},
  \href{http://arxiv.org/abs/1704.07410}{{\ttfamily arXiv:1704.07410
  [hep-th]}}.

\bibitem{Suzuki:2021zbe}
K.~Suzuki and T.~Takayanagi, ``{JT gravity limit of Liouville CFT and matrix
  model},'' \href{http://dx.doi.org/10.1007/JHEP11(2021)137}{{\em JHEP}
  {\bfseries 11} (2021) 137}, \href{http://arxiv.org/abs/2108.12096}{{\ttfamily
  arXiv:2108.12096 [hep-th]}}.

\bibitem{Goel:2020yxl}
A.~Goel, L.~V. Iliesiu, J.~Kruthoff, and Z.~Yang, ``{Classifying boundary
  conditions in JT gravity: from energy-branes to $\alpha$-branes},''
  \href{http://dx.doi.org/10.1007/JHEP04(2021)069}{{\em JHEP} {\bfseries 04}
  (2021) 069}, \href{http://arxiv.org/abs/2010.12592}{{\ttfamily
  arXiv:2010.12592 [hep-th]}}.

\bibitem{Collier:2023cyw}
S.~Collier, L.~Eberhardt, B.~M\"uhlmann, and V.~A. Rodriguez, ``{The Virasoro
  Minimal String},'' \href{http://arxiv.org/abs/2309.10846}{{\ttfamily
  arXiv:2309.10846 [hep-th]}}.

\bibitem{Ikeda:1993aj}
N.~Ikeda and K.~I. Izawa, ``{General form of dilaton gravity and nonlinear
  gauge theory},'' \href{http://dx.doi.org/10.1143/PTP.90.237}{{\em Prog.
  Theor. Phys.} {\bfseries 90} (1993) 237--246},
  \href{http://arxiv.org/abs/hep-th/9304012}{{\ttfamily arXiv:hep-th/9304012}}.

\bibitem{Ikeda:1993fh}
N.~Ikeda, ``{Two-dimensional gravity and nonlinear gauge theory},''
  \href{http://dx.doi.org/10.1006/aphy.1994.1104}{{\em Annals Phys.} {\bfseries
  235} (1994) 435--464}, \href{http://arxiv.org/abs/hep-th/9312059}{{\ttfamily
  arXiv:hep-th/9312059}}.

\bibitem{Schaller:1994es}
P.~Schaller and T.~Strobl, ``{Poisson structure induced (topological) field
  theories},'' \href{http://dx.doi.org/10.1142/S0217732394002951}{{\em Mod.
  Phys. Lett. A} {\bfseries 9} (1994) 3129--3136},
  \href{http://arxiv.org/abs/hep-th/9405110}{{\ttfamily arXiv:hep-th/9405110}}.

\bibitem{Blommaert:2018oro}
A.~Blommaert, T.~G. Mertens, and H.~Verschelde, ``{The Schwarzian Theory - A
  Wilson Line Perspective},''
  \href{http://dx.doi.org/10.1007/JHEP12(2018)022}{{\em JHEP} {\bfseries 12}
  (2018) 022}, \href{http://arxiv.org/abs/1806.07765}{{\ttfamily
  arXiv:1806.07765 [hep-th]}}.

\bibitem{Iliesiu:2019xuh}
L.~V. Iliesiu, S.~S. Pufu, H.~Verlinde, and Y.~Wang, ``{An exact quantization
  of Jackiw-Teitelboim gravity},''
  \href{http://dx.doi.org/10.1007/JHEP11(2019)091}{{\em JHEP} {\bfseries 11}
  (2019) 091}, \href{http://arxiv.org/abs/1905.02726}{{\ttfamily
  arXiv:1905.02726 [hep-th]}}.

\bibitem{polyakov1981quantum}
A.~M. Polyakov, ``Quantum geometry of bosonic strings,'' {\em Physics Letters
  B} {\bfseries 103} no.~3, (1981) 207--210.

\bibitem{david1988conformal}
F.~David, ``Conformal field theories coupled to 2-d gravity in the conformal
  gauge,'' {\em Modern Physics Letters A} {\bfseries 3} no.~17, (1988)
  1651--1656.

\bibitem{distler1989conformal}
J.~Distler and H.~Kawai, ``Conformal field theory and 2d quantum gravity,''
  {\em Nuclear physics B} {\bfseries 321} no.~2, (1989) 509--527.

\bibitem{StanfordSeiberg}
D.~Stanford and N.~Seiberg, ``{unpublished},'' 2019.

\bibitem{Turiaci:2020fjj}
G.~J. Turiaci, M.~Usatyuk, and W.~W. Weng, ``{2D dilaton-gravity, deformations
  of the minimal string, and matrix models},''
  \href{http://dx.doi.org/10.1088/1361-6382/ac25df}{{\em Class. Quant. Grav.}
  {\bfseries 38} no.~20, (2021) 204001},
  \href{http://arxiv.org/abs/2011.06038}{{\ttfamily arXiv:2011.06038
  [hep-th]}}.

\bibitem{Fateev:2000ik}
V.~Fateev, A.~B. Zamolodchikov, and A.~B. Zamolodchikov, ``{Boundary Liouville
  field theory. 1. Boundary state and boundary two point function},''
  \href{http://arxiv.org/abs/hep-th/0001012}{{\ttfamily arXiv:hep-th/0001012}}.

\bibitem{Teschner:2001rv}
J.~Teschner, ``{Liouville theory revisited},''
  \href{http://dx.doi.org/10.1088/0264-9381/18/23/201}{{\em Class. Quant.
  Grav.} {\bfseries 18} (2001) R153--R222},
  \href{http://arxiv.org/abs/hep-th/0104158}{{\ttfamily arXiv:hep-th/0104158}}.

\bibitem{Narovlansky:2023lfz}
V.~Narovlansky and H.~Verlinde, ``{Double-scaled SYK and de Sitter
  Holography},'' \href{http://arxiv.org/abs/2310.16994}{{\ttfamily
  arXiv:2310.16994 [hep-th]}}.

\bibitem{Berkooz:2022mfk}
M.~Berkooz, M.~Isachenkov, P.~Narayan, and V.~Narovlansky, ``{Quantum groups,
  non-commutative $AdS_2$, and chords in the double-scaled SYK model},''
  \href{http://arxiv.org/abs/2212.13668}{{\ttfamily arXiv:2212.13668
  [hep-th]}}.

\bibitem{Mertens:2022aou}
T.~G. Mertens, ``{Quantum exponentials for the modular double and applications
  in gravity models},'' \href{http://dx.doi.org/10.1007/JHEP09(2023)106}{{\em
  JHEP} {\bfseries 09} (2023) 106},
  \href{http://arxiv.org/abs/2212.07696}{{\ttfamily arXiv:2212.07696
  [hep-th]}}.

\bibitem{Faddeev:1999fe}
L.~D. Faddeev, ``{Modular double of quantum group},'' in {\em {Conference Moshe
  Flato}}, pp.~149--156.
\newblock 2000.
\newblock \href{http://arxiv.org/abs/math/9912078}{{\ttfamily
  arXiv:math/9912078}}.

\bibitem{Ponsot:1999uf}
B.~Ponsot and J.~Teschner, ``{Liouville bootstrap via harmonic analysis on a
  noncompact quantum group},''
  \href{http://arxiv.org/abs/hep-th/9911110}{{\ttfamily arXiv:hep-th/9911110}}.

\bibitem{Ponsot:2000mt}
B.~Ponsot and J.~Teschner, ``{Clebsch-Gordan and Racah-Wigner coefficients for
  a continuous series of representations of U(q)(sl(2,R))},''
  \href{http://dx.doi.org/10.1007/PL00005590}{{\em Commun. Math. Phys.}
  {\bfseries 224} (2001) 613--655},
  \href{http://arxiv.org/abs/math/0007097}{{\ttfamily arXiv:math/0007097}}.

\bibitem{Kharchev:2001rs}
S.~Kharchev, D.~Lebedev, and M.~Semenov-Tian-Shansky, ``{Unitary
  representations of U(q) (sl(2, R)), the modular double, and the multiparticle
  q deformed Toda chains},''
  \href{http://dx.doi.org/10.1007/s002200100592}{{\em Commun. Math. Phys.}
  {\bfseries 225} (2002) 573--609},
  \href{http://arxiv.org/abs/hep-th/0102180}{{\ttfamily arXiv:hep-th/0102180}}.

\bibitem{Bytsko:2002br}
A.~G. Bytsko and J.~Teschner, ``{R operator, coproduct and Haar measure for the
  modular double of U(q)(sl(2,R))},''
  \href{http://dx.doi.org/10.1007/s00220-003-0894-5}{{\em Commun. Math. Phys.}
  {\bfseries 240} (2003) 171--196},
\href{http://arxiv.org/abs/math/0208191}{{\ttfamily arXiv:math/0208191
  [math-qa]}}.

\bibitem{Bytsko:2006ut}
A.~G. Bytsko and J.~Teschner, ``{Quantization of models with non-compact
  quantum group symmetry: Modular XXZ magnet and lattice sinh-Gordon model},''
  \href{http://dx.doi.org/10.1088/0305-4470/39/41/S11}{{\em J. Phys.}
  {\bfseries A39} (2006) 12927--12981},
\href{http://arxiv.org/abs/hep-th/0602093}{{\ttfamily arXiv:hep-th/0602093
  [hep-th]}}.

\bibitem{klimyk2012quantum}
A.~Klimyk and K.~Schm{\"u}dgen, {\em Quantum groups and their representations}.
\newblock Springer Science \& Business Media, 2012.

\bibitem{Bagrets:2016cdf}
D.~Bagrets, A.~Altland, and A.~Kamenev,
  ``{Sachdev\textendash{}Ye\textendash{}Kitaev model as Liouville quantum
  mechanics},'' \href{http://dx.doi.org/10.1016/j.nuclphysb.2016.08.002}{{\em
  Nucl. Phys. B} {\bfseries 911} (2016) 191--205},
  \href{http://arxiv.org/abs/1607.00694}{{\ttfamily arXiv:1607.00694
  [cond-mat.str-el]}}.

\bibitem{Harlow:2018tqv}
D.~Harlow and D.~Jafferis, ``{The Factorization Problem in Jackiw-Teitelboim
  Gravity},'' \href{http://dx.doi.org/10.1007/JHEP02(2020)177}{{\em JHEP}
  {\bfseries 02} (2020) 177}, \href{http://arxiv.org/abs/1804.01081}{{\ttfamily
  arXiv:1804.01081 [hep-th]}}.

\bibitem{brown1986central}
J.~D. Brown and M.~Henneaux, ``Central charges in the canonical realization of
  asymptotic symmetries: an example from three dimensional gravity,'' {\em
  Communications in Mathematical Physics} {\bfseries 104} (1986) 207--226.

\bibitem{Maldacena:2016hyu}
J.~Maldacena and D.~Stanford, ``{Remarks on the Sachdev-Ye-Kitaev model},''
  \href{http://dx.doi.org/10.1103/PhysRevD.94.106002}{{\em Phys. Rev. D}
  {\bfseries 94} no.~10, (2016) 106002},
  \href{http://arxiv.org/abs/1604.07818}{{\ttfamily arXiv:1604.07818
  [hep-th]}}.

\bibitem{Lin:2023trc}
H.~W. Lin and D.~Stanford, ``{A symmetry algebra in double-scaled SYK},''
  \href{http://arxiv.org/abs/2307.15725}{{\ttfamily arXiv:2307.15725
  [hep-th]}}.

\bibitem{Lin:2022rbf}
H.~W. Lin, ``{The bulk Hilbert space of double scaled SYK},''
  \href{http://dx.doi.org/10.1007/JHEP11(2022)060}{{\em JHEP} {\bfseries 11}
  (2022) 060}, \href{http://arxiv.org/abs/2208.07032}{{\ttfamily
  arXiv:2208.07032 [hep-th]}}.

\bibitem{Haake:1315494}
F.~Haake, S.~Gnutzmann, and M.~Kuś,
  \href{http://dx.doi.org/10.1007/978-3-642-05428-0}{{\em {Quantum Signatures
  of Chaos; 4th ed.}}}
\newblock Springer series in synergetics. Springer, Dordrecht, 2018.

\bibitem{Seiberg:2003nm}
N.~Seiberg and D.~Shih, ``{Branes, rings and matrix models in minimal
  (super)string theory},''
  \href{http://dx.doi.org/10.1088/1126-6708/2004/02/021}{{\em JHEP} {\bfseries
  02} (2004) 021}, \href{http://arxiv.org/abs/hep-th/0312170}{{\ttfamily
  arXiv:hep-th/0312170}}.

\bibitem{Saad:2019lba}
P.~Saad, S.~H. Shenker, and D.~Stanford, ``{JT gravity as a matrix integral},''
  \href{http://arxiv.org/abs/1903.11115}{{\ttfamily arXiv:1903.11115
  [hep-th]}}.

\bibitem{ip2021tensor}
I.~C. Ip, ``On tensor product decomposition of positive representations of
  u$_q(sl(2, r))$,'' {\em Letters in Mathematical Physics} {\bfseries 111}
  (2021) 1--35, \href{http://arxiv.org/abs/1511.07970}{{\ttfamily
  arXiv:1511.07970 [math]}}.

\bibitem{Gegenberg:1994pv}
J.~Gegenberg, G.~Kunstatter, and D.~Louis-Martinez, ``{Observables for
  two-dimensional black holes},''
  \href{http://dx.doi.org/10.1103/PhysRevD.51.1781}{{\em Phys. Rev. D}
  {\bfseries 51} (1995) 1781--1786},
  \href{http://arxiv.org/abs/gr-qc/9408015}{{\ttfamily arXiv:gr-qc/9408015}}.

\bibitem{Witten:2020ert}
E.~Witten, ``{Deformations of JT Gravity and Phase Transitions},''
  \href{http://arxiv.org/abs/2006.03494}{{\ttfamily arXiv:2006.03494
  [hep-th]}}.

\bibitem{Carlip:1993sa}
S.~Carlip and C.~Teitelboim, ``{The Off-shell black hole},''
  \href{http://dx.doi.org/10.1088/0264-9381/12/7/011}{{\em Class. Quant. Grav.}
  {\bfseries 12} (1995) 1699--1704},
  \href{http://arxiv.org/abs/gr-qc/9312002}{{\ttfamily arXiv:gr-qc/9312002}}.

\bibitem{Susskind:1994sm}
L.~Susskind and J.~Uglum, ``{Black hole entropy in canonical quantum gravity
  and superstring theory},''
  \href{http://dx.doi.org/10.1103/PhysRevD.50.2700}{{\em Phys. Rev. D}
  {\bfseries 50} (1994) 2700--2711},
  \href{http://arxiv.org/abs/hep-th/9401070}{{\ttfamily arXiv:hep-th/9401070}}.

\bibitem{Dong:2018seb}
X.~Dong, D.~Harlow, and D.~Marolf, ``{Flat entanglement spectra in fixed-area
  states of quantum gravity},''
  \href{http://dx.doi.org/10.1007/JHEP10(2019)240}{{\em JHEP} {\bfseries 10}
  (2019) 240}, \href{http://arxiv.org/abs/1811.05382}{{\ttfamily
  arXiv:1811.05382 [hep-th]}}.

\bibitem{Dong:2022ilf}
X.~Dong, D.~Marolf, P.~Rath, A.~Tajdini, and Z.~Wang, ``{The spacetime geometry
  of fixed-area states in gravitational systems},''
  \href{http://dx.doi.org/10.1007/JHEP08(2022)158}{{\em JHEP} {\bfseries 08}
  (2022) 158}, \href{http://arxiv.org/abs/2203.04973}{{\ttfamily
  arXiv:2203.04973 [hep-th]}}.

\bibitem{henneaux1985quantum}
M.~Henneaux, ``Quantum gravity in two dimensions: Exact solution of the jackiw
  model,'' {\em Physical review letters} {\bfseries 54} no.~10, (1985) 959.

\bibitem{Louis-Martinez:1993bge}
D.~Louis-Martinez, J.~Gegenberg, and G.~Kunstatter, ``{Exact Dirac quantization
  of all 2-D dilaton gravity theories},''
  \href{http://dx.doi.org/10.1016/0370-2693(94)90463-4}{{\em Phys. Lett. B}
  {\bfseries 321} (1994) 193--198},
  \href{http://arxiv.org/abs/gr-qc/9309018}{{\ttfamily arXiv:gr-qc/9309018}}.

\bibitem{Iliesiu:2020zld}
L.~V. Iliesiu, J.~Kruthoff, G.~J. Turiaci, and H.~Verlinde, ``{JT gravity at
  finite cutoff},'' \href{http://dx.doi.org/10.21468/SciPostPhys.9.2.023}{{\em
  SciPost Phys.} {\bfseries 9} (2020) 023},
  \href{http://arxiv.org/abs/2004.07242}{{\ttfamily arXiv:2004.07242
  [hep-th]}}.

\bibitem{Blommaert:2023vbz}
A.~Blommaert, J.~Kruthoff, and S.~Yao, ``{The power of Lorentzian wormholes},''
  \href{http://arxiv.org/abs/2302.01360}{{\ttfamily arXiv:2302.01360
  [hep-th]}}.

\bibitem{knizhnik1988fractal}
V.~G. Knizhnik, A.~M. Polyakov, and A.~B. Zamolodchikov, ``Fractal structure of
  2d—quantum gravity,'' {\em Modern Physics Letters A} {\bfseries 3} no.~08,
  (1988) 819--826.

\bibitem{Yang:2018gdb}
Z.~Yang, ``{The Quantum Gravity Dynamics of Near Extremal Black Holes},''
  \href{http://dx.doi.org/10.1007/JHEP05(2019)205}{{\em JHEP} {\bfseries 05}
  (2019) 205}, \href{http://arxiv.org/abs/1809.08647}{{\ttfamily
  arXiv:1809.08647 [hep-th]}}.

\bibitem{Saad:2019pqd}
P.~Saad, ``{Late Time Correlation Functions, Baby Universes, and ETH in JT
  Gravity},'' \href{http://arxiv.org/abs/1910.10311}{{\ttfamily
  arXiv:1910.10311 [hep-th]}}.

\bibitem{Lin:2022zxd}
H.~W. Lin, J.~Maldacena, L.~Rozenberg, and J.~Shan, ``{Looking at
  supersymmetric black holes for a very long time},''
  \href{http://dx.doi.org/10.21468/SciPostPhys.14.5.128}{{\em SciPost Phys.}
  {\bfseries 14} no.~5, (2023) 128},
  \href{http://arxiv.org/abs/2207.00408}{{\ttfamily arXiv:2207.00408
  [hep-th]}}.

\bibitem{Goel:2023svz}
A.~Goel, V.~Narovlansky, and H.~Verlinde, ``{Semiclassical geometry in
  double-scaled SYK},'' \href{http://arxiv.org/abs/2301.05732}{{\ttfamily
  arXiv:2301.05732 [hep-th]}}.

\bibitem{Lam:2018pvp}
H.~T. Lam, T.~G. Mertens, G.~J. Turiaci, and H.~Verlinde, ``{Shockwave S-matrix
  from Schwarzian Quantum Mechanics},''
  \href{http://dx.doi.org/10.1007/JHEP11(2018)182}{{\em JHEP} {\bfseries 11}
  (2018) 182}, \href{http://arxiv.org/abs/1804.09834}{{\ttfamily
  arXiv:1804.09834 [hep-th]}}.

\bibitem{Shenker:2014cwa}
S.~H. Shenker and D.~Stanford, ``{Stringy effects in scrambling},''
  \href{http://dx.doi.org/10.1007/JHEP05(2015)132}{{\em JHEP} {\bfseries 05}
  (2015) 132}, \href{http://arxiv.org/abs/1412.6087}{{\ttfamily arXiv:1412.6087
  [hep-th]}}.

\bibitem{jaganathan2000introduction}
R.~Jaganathan, ``An introduction to quantum algebras and their applications,''
  {\em arXiv preprint math-ph/0003018} (2000) .

\bibitem{Belaey:2023jtr}
A.~Belaey, F.~Mariani, and T.~G. Mertens, ``{Branes in JT (super)gravity from
  group theory},'' \href{http://arxiv.org/abs/2310.04245}{{\ttfamily
  arXiv:2310.04245 [hep-th]}}.

\bibitem{Blommaert:2022ucs}
A.~Blommaert, L.~V. Iliesiu, and J.~Kruthoff, ``{Alpha states demystified
  \textemdash{} towards microscopic models of AdS$_{2}$ holography},''
  \href{http://dx.doi.org/10.1007/JHEP08(2022)071}{{\em JHEP} {\bfseries 08}
  (2022) 071}, \href{http://arxiv.org/abs/2203.07384}{{\ttfamily
  arXiv:2203.07384 [hep-th]}}.

\bibitem{Cattaneo:2001bp}
A.~S. Cattaneo and G.~Felder, ``{Poisson sigma models and deformation
  quantization},'' \href{http://dx.doi.org/10.1142/S0217732301003255}{{\em Mod.
  Phys. Lett. A} {\bfseries 16} (2001) 179--190},
  \href{http://arxiv.org/abs/hep-th/0102208}{{\ttfamily arXiv:hep-th/0102208}}.

\bibitem{Mertens:2018fds}
T.~G. Mertens, ``{The Schwarzian theory \textemdash{} origins},''
  \href{http://dx.doi.org/10.1007/JHEP05(2018)036}{{\em JHEP} {\bfseries 05}
  (2018) 036}, \href{http://arxiv.org/abs/1801.09605}{{\ttfamily
  arXiv:1801.09605 [hep-th]}}.

\bibitem{Blommaert:2018oue}
A.~Blommaert, T.~G. Mertens, and H.~Verschelde, ``{Edge dynamics from the path
  integral \textemdash{} Maxwell and Yang-Mills},''
  \href{http://dx.doi.org/10.1007/JHEP11(2018)080}{{\em JHEP} {\bfseries 11}
  (2018) 080}, \href{http://arxiv.org/abs/1804.07585}{{\ttfamily
  arXiv:1804.07585 [hep-th]}}.

\bibitem{Blommaert:2018iqz}
A.~Blommaert, T.~G. Mertens, and H.~Verschelde, ``{Fine Structure of
  Jackiw-Teitelboim Quantum Gravity},''
  \href{http://dx.doi.org/10.1007/JHEP09(2019)066}{{\em JHEP} {\bfseries 09}
  (2019) 066}, \href{http://arxiv.org/abs/1812.00918}{{\ttfamily
  arXiv:1812.00918 [hep-th]}}.

\bibitem{Migdal:1975zg}
A.~A. Migdal, ``{Recursion Equations in Gauge Theories},'' {\em Sov. Phys.
  JETP} {\bfseries 42} (1975) 413.

\bibitem{Witten:1992xu}
E.~Witten, ``{Two-dimensional gauge theories revisited},''
  \href{http://dx.doi.org/10.1016/0393-0440(92)90034-X}{{\em J. Geom. Phys.}
  {\bfseries 9} (1992) 303--368},
  \href{http://arxiv.org/abs/hep-th/9204083}{{\ttfamily arXiv:hep-th/9204083}}.

\bibitem{Cordes:1994fc}
S.~Cordes, G.~W. Moore, and S.~Ramgoolam, ``{Lectures on 2-d Yang-Mills theory,
  equivariant cohomology and topological field theories},''
  \href{http://dx.doi.org/10.1016/0920-5632(95)00434-B}{{\em Nucl. Phys. B
  Proc. Suppl.} {\bfseries 41} (1995) 184--244},
  \href{http://arxiv.org/abs/hep-th/9411210}{{\ttfamily arXiv:hep-th/9411210}}.

\bibitem{Witten:1988hc}
E.~Witten, ``{(2+1)-Dimensional Gravity as an Exactly Soluble System},''
  \href{http://dx.doi.org/10.1016/0550-3213(88)90143-5}{{\em Nucl. Phys. B}
  {\bfseries 311} (1988) 46}.

\bibitem{Witten:2007kt}
E.~Witten, ``{Three-Dimensional Gravity Revisited},''
  \href{http://arxiv.org/abs/0706.3359}{{\ttfamily arXiv:0706.3359 [hep-th]}}.

\bibitem{Donnay:2016iyk}
L.~Donnay, ``{Asymptotic dynamics of three-dimensional gravity},''
  \href{http://dx.doi.org/10.22323/1.271.0001}{{\em PoS} {\bfseries Modave2015}
  (2016) 001}, \href{http://arxiv.org/abs/1602.09021}{{\ttfamily
  arXiv:1602.09021 [hep-th]}}.

\bibitem{Cotler:2020ugk}
J.~Cotler and K.~Jensen, ``{AdS$_{3}$ gravity and random CFT},''
  \href{http://dx.doi.org/10.1007/JHEP04(2021)033}{{\em JHEP} {\bfseries 04}
  (2021) 033}, \href{http://arxiv.org/abs/2006.08648}{{\ttfamily
  arXiv:2006.08648 [hep-th]}}.

\bibitem{Collier:2023fwi}
S.~Collier, L.~Eberhardt, and M.~Zhang, ``{Solving 3d Gravity with Virasoro
  TQFT},'' \href{http://arxiv.org/abs/2304.13650}{{\ttfamily arXiv:2304.13650
  [hep-th]}}.

\bibitem{Fukuyama:1985gg}
T.~Fukuyama and K.~Kamimura, ``{Gauge Theory of Two-dimensional Gravity},''
  \href{http://dx.doi.org/10.1016/0370-2693(85)91322-X}{{\em Phys. Lett. B}
  {\bfseries 160} (1985) 259--262}.

\bibitem{Isler:1989hq}
K.~Isler and C.~A. Trugenberger, ``{A Gauge Theory of Two-dimensional Quantum
  Gravity},'' \href{http://dx.doi.org/10.1103/PhysRevLett.63.834}{{\em Phys.
  Rev. Lett.} {\bfseries 63} (1989) 834}.

\bibitem{Chamseddine:1989yz}
A.~H. Chamseddine and D.~Wyler, ``{Gauge Theory of Topological Gravity in
  (1+1)-Dimensions},''
  \href{http://dx.doi.org/10.1016/0370-2693(89)90528-5}{{\em Phys. Lett. B}
  {\bfseries 228} (1989) 75--78}.

\bibitem{Jackiw:1992bw}
R.~Jackiw, ``{Gauge theories for gravity on a line},''
  \href{http://dx.doi.org/10.1007/BF01017075}{{\em Theor. Math. Phys.}
  {\bfseries 92} (1992) 979--987},
  \href{http://arxiv.org/abs/hep-th/9206093}{{\ttfamily arXiv:hep-th/9206093}}.

\bibitem{Grumiller:2021cwg}
D.~Grumiller, R.~Ruzziconi, and C.~Zwikel, ``{Generalized dilaton gravity in
  2d},'' \href{http://dx.doi.org/10.21468/SciPostPhys.12.1.032}{{\em SciPost
  Phys.} {\bfseries 12} no.~1, (2022) 032},
  \href{http://arxiv.org/abs/2109.03266}{{\ttfamily arXiv:2109.03266
  [hep-th]}}.

\bibitem{Grumiller:2020elf}
D.~Grumiller, J.~Hartong, S.~Prohazka, and J.~Salzer, ``{Limits of JT
  gravity},'' \href{http://dx.doi.org/10.1007/JHEP02(2021)134}{{\em JHEP}
  {\bfseries 02} (2021) 134}, \href{http://arxiv.org/abs/2011.13870}{{\ttfamily
  arXiv:2011.13870 [hep-th]}}.

\bibitem{Klosch:1995fi}
T.~Klosch and T.~Strobl, ``{Classical and quantum gravity in (1+1)-Dimensions.
  Part 1: A Unifying approach},''
  \href{http://dx.doi.org/10.1088/0264-9381/13/5/015}{{\em Class. Quant. Grav.}
  {\bfseries 13} (1996) 965--984},
  \href{http://arxiv.org/abs/gr-qc/9508020}{{\ttfamily arXiv:gr-qc/9508020}}.
  [Erratum: Class.Quant.Grav. 14, 825 (1997)].

\bibitem{Maxfield:2020ale}
H.~Maxfield and G.~J. Turiaci, ``{The path integral of 3D gravity near
  extremality; or, JT gravity with defects as a matrix integral},''
  \href{http://dx.doi.org/10.1007/JHEP01(2021)118}{{\em JHEP} {\bfseries 01}
  (2021) 118}, \href{http://arxiv.org/abs/2006.11317}{{\ttfamily
  arXiv:2006.11317 [hep-th]}}.

\bibitem{Witten:2020wvy}
E.~Witten, ``{Matrix Models and Deformations of JT Gravity},''
  \href{http://dx.doi.org/10.1098/rspa.2020.0582}{{\em Proc. Roy. Soc. Lond. A}
  {\bfseries 476} no.~2244, (2020) 20200582},
  \href{http://arxiv.org/abs/2006.13414}{{\ttfamily arXiv:2006.13414
  [hep-th]}}.

\bibitem{Stanford:2020wkf}
D.~Stanford, ``{More quantum noise from wormholes},''
  \href{http://arxiv.org/abs/2008.08570}{{\ttfamily arXiv:2008.08570
  [hep-th]}}.

\bibitem{Lin:2022nss}
H.~Lin and L.~Susskind, ``{Infinite Temperature's Not So Hot},''
  \href{http://arxiv.org/abs/2206.01083}{{\ttfamily arXiv:2206.01083
  [hep-th]}}.

\bibitem{Mertens:2019tcm}
T.~G. Mertens and G.~J. Turiaci, ``{Defects in Jackiw-Teitelboim Quantum
  Gravity},'' \href{http://dx.doi.org/10.1007/JHEP08(2019)127}{{\em JHEP}
  {\bfseries 08} (2019) 127}, \href{http://arxiv.org/abs/1904.05228}{{\ttfamily
  arXiv:1904.05228 [hep-th]}}.

\bibitem{Mertens:2022ujr}
T.~G. Mertens, J.~Sim\'on, and G.~Wong, ``{A proposal for 3d quantum gravity
  and its bulk factorization},''
  \href{http://arxiv.org/abs/2210.14196}{{\ttfamily arXiv:2210.14196
  [hep-th]}}.

\bibitem{Wong:2022eiu}
G.~Wong, ``{A note on the bulk interpretation of the Quantum Extremal Surface
  formula},'' \href{http://arxiv.org/abs/2212.03193}{{\ttfamily
  arXiv:2212.03193 [hep-th]}}.

\bibitem{Felder:1999mq}
G.~Felder, J.~Frohlich, J.~Fuchs, and C.~Schweigert, ``{Correlation functions
  and boundary conditions in RCFT and three-dimensional topology},''
  \href{http://dx.doi.org/10.1023/A:1014903315415}{{\em Compos. Math.}
  {\bfseries 131} (2002) 189--237},
  \href{http://arxiv.org/abs/hep-th/9912239}{{\ttfamily arXiv:hep-th/9912239}}.

\bibitem{Iliesiu:2020qvm}
L.~V. Iliesiu and G.~J. Turiaci, ``{The statistical mechanics of near-extremal
  black holes},'' \href{http://dx.doi.org/10.1007/JHEP05(2021)145}{{\em JHEP}
  {\bfseries 05} (2021) 145}, \href{http://arxiv.org/abs/2003.02860}{{\ttfamily
  arXiv:2003.02860 [hep-th]}}.

\bibitem{Almheiri:2016fws}
A.~Almheiri and B.~Kang, ``{Conformal Symmetry Breaking and Thermodynamics of
  Near-Extremal Black Holes},''
  \href{http://dx.doi.org/10.1007/JHEP10(2016)052}{{\em JHEP} {\bfseries 10}
  (2016) 052}, \href{http://arxiv.org/abs/1606.04108}{{\ttfamily
  arXiv:1606.04108 [hep-th]}}.

\bibitem{Nayak:2018qej}
P.~Nayak, A.~Shukla, R.~M. Soni, S.~P. Trivedi, and V.~Vishal, ``{On the
  Dynamics of Near-Extremal Black Holes},''
  \href{http://dx.doi.org/10.1007/JHEP09(2018)048}{{\em JHEP} {\bfseries 09}
  (2018) 048}, \href{http://arxiv.org/abs/1802.09547}{{\ttfamily
  arXiv:1802.09547 [hep-th]}}.

\bibitem{Moitra:2018jqs}
U.~Moitra, S.~P. Trivedi, and V.~Vishal, ``{Extremal and near-extremal black
  holes and near-CFT$_{1}$},''
  \href{http://dx.doi.org/10.1007/JHEP07(2019)055}{{\em JHEP} {\bfseries 07}
  (2019) 055}, \href{http://arxiv.org/abs/1808.08239}{{\ttfamily
  arXiv:1808.08239 [hep-th]}}.

\end{thebibliography}\endgroup

\end{document}